\documentclass{aa}  
\usepackage{graphicx,color}
\usepackage{txfonts}
\usepackage{hyperref}
\bibpunct{(}{)}{;}{a}{}{,} 

\newcommand{\rosat}{{\it ROSAT~}}

\newcommand{\itg}{{\it INTEGRAL~}}
\newcommand{\itgu}{{\it INTEGRAL}}
\newcommand{\gaia}{{\it Gaia~}}
\newcommand{\hip}{{\it HIPPARCOS~}}
\newcommand{\her}{{\it HERSCHEL~}}
\newcommand{\heru}{{\it HERSCHEL}}

\newcommand{\apexu}{{\it APEX}}
\newcommand{\planck}{{\it PLANCK~}}
\newcommand{\hi}[1]{#1}
\newcommand{\al}{{$^{26}$Al~}}
\newcommand{\alu}{{$^{26}$Al}}
\newcommand{\HI}{H{\sc{i}}~}
\newcommand{\NaI}{ Na{\sc{i}}~}
\newcommand{\msun}{{M$_\odot$}}

\newcommand{\kms}{{km~s$^{-1}$}}
\newcommand{\sbr}[1]{\left[ #1 \right]} 

\renewcommand{\deg}{^\circ}

\newcommand{\rvb}[1]{{ #1}}

\begin{document}

\title{Surround \& Squash: The Interstellar Medium around \\ Scorpius Centaurus OB2}
\titlerunning{Surround \& Squash in Sco-Cen}
 \author{Martin G. H. Krause \inst{1,2,3,4,5} \fnmsep\thanks{E-mail:
    M.G.H.Krause@herts.ac.uk} 
   \and Andreas Burkert \inst{1,4,5}\fnmsep\thanks{Max-Planck-Fellow} 
	\and Roland Diehl \inst{4,5} 
\and  Katharina Fierlinger\inst{1,5}\\
\and Benjamin Gaczkowski \inst{1}
\and Daniel Kroell \inst{1,4}
\and  Judith Ngoumou\inst{1}
\and Veronica Roccatagliata \inst{6,1}
\and Thomas Siegert \inst{4}
\and Thomas Preibisch \inst{1,5}
}

\institute{
Universit\"ats-Sternwarte M\"unchen, Ludwig-Maximilians-Universit\"at,
Scheinerstr. 1, 81679 M\"unchen, Germany
\and  School of Physical Sciences, University of Tasmania, Hobart, TAS, 7005, Australia 
\and Centre for Astrophysics Research, School of Physics, Astronomy and Mathematics, University of Hertfordshire, College Lane, Hatfield, Hertfordshire AL10 9AB, UK
\and Max-Planck-Institut f\"ur extraterrestrische Physik,
  Giessenbachstr.~1, 85741 Garching, Germany 
\and Excellence Cluster Universe, Technische Universit\"at
  M\"unchen, Boltzmannstrasse 2, 85748 Garching, Germany 
\and INAF -- Osservatorio Astrofisico di Arcetri,
Largo E. Fermi 5, 50125, Firenze, Italy
}

   \date{4 Dec 2017; received / accepted ...}

 
  \abstract
   {
	\hi{Feedback by massive stars shapes the interstellar medium and is thought to influence subsequent star
	formation. Details of this process are under debate.}
   	 }
   {
	\hi{We exploited observational constraints on stars, gas and nucleosynthesis ashes for the closest region
	of recent massive-star formation, Scorpius-Centaurus OB2, and combined 
	them with 3D hydrodynamical simulations, in order to address physics and history for the case of  
	the Scorpius-Centaurus superbubble.}}
   {\hi{We used published cold 
	gas observations through \planck survey data processing, \her and 
	\apexu, continuum and molecular line observations. We analysed the Galactic All Sky Survey (GASS) 
	to investigate shell structures in atomic hydrogen, and used \hip and \gaia data in combination with
	interstellar absorption against stars to obtain new constraints for the distance to the \HI features. Hot gas is traced in soft X-rays via 
	the \rosat all sky survey. Nucleosynthesis ejecta from massive stars were traced with new \itg
	spectrometer observations via $^{26}$Al radioactivity.
	We also performed 3D hydrodynamical simulations for the Sco-Cen superbubble. }
}
   { \hi{Soft X-rays and  a now more significant 
   	detection of $^{26}$Al confirm recent ($\approx 1$~Myr ago) input of mass, energy and 
   	nucleosynthesis ejecta, likely by a supernova in the Upper Scorpius (USco) subgroup. 
   	We confirm a large supershell around the entire 
   	OB association and perform a 3D hydrodynamics simulations with a conservative massive star population that
   	reproduces the morphology of the superbubble.  High resolution GASS observations of a nested 
   	supershell reveal
   	that it is filamentary \rvb{possibly related to} the Vishniac clumping instability, but molecular gas (Lupus~I) 
   	is only present where the shell coincides
   	with the connecting line between the subgroups of the OB association, suggesting
   	a connection to the cloud, probably an elongated sheet, out of 
   	which the OB association formed. 
	Stars have formed sequentially in the subgroups of the OB association and currently form in Lupus~I.
	To investigate the impact of massive star feedback on extended clouds, 
	we simulate the interaction of a turbulent cloud with the hot, pressurised gas in a superbubble.
	The hot gas fills the tenuous regions of the cloud and compresses the denser parts.
	Stars formed in these dense clumps would have distinct spatial and kinematic distributions.}
	}
   {\hi{The combined results from observations and simulations are consistent with a scenario where dense gas was initially distributed
	in a band elongated in the direction now occupied by the OB association.  
	Superbubbles powered by massive stars would then repeatedly break out of the
	elongated parent cloud, surround and squash the denser parts of the gas sheet and thus induce
	more star formation. The expected spatial and kinematic distribution of stars is consistent with
	observations of Sco-Cen.
	The scenario might apply to many similar regions in the Galaxy and also to AGN-related superbubbles.}
	 }

   \maketitle

\section{Introduction}\label{sec:i} 
The interstellar medium is a multifaceted, dynamic place:
massive stars inject energy, mass and freshly synthesised nuclei. This changes its metallicity and 
sustains multi-phase turbulence with volume-filling, hot and rarefied regions and
dense layers compressed by converging flows which may make the transition to the molecular
phase and form the next generation of stars
\citep{MO77,dAB05,Heitschea08,Hennea08,Micicea13,GomVaz14,GongOst15a}. Key drivers of this 
process are massive stars with their ionising radiation, winds
and core-collapse supernovae.

It is clear that in order to understand these complex phase transitions in the interstellar medium
associated with multiple generations of stars, studies in any single wavelength cannot be sufficient.
Molecular gas, often associated with ongoing star formation, is best studied at infrared and sub-mm 
wavelengths. Atomic gas is connected with shells pushed out by energy input of young massive stars
and efficiently mapped in the 21~cm radio line and with optical absorption against stars. 
Tenuous gas shock-heated by the winds and explosions of massive stars to keV temperatures
is only seen in X-ray observations, and the ejecta themselves can be traced via their radioactive 
decay lines at MeV energies.

\begin{table*}
\caption{Sco-Cen subgroups and their properties}
\label{t:sc-overview}      
\centering                          
\begin{tabular}{ccccccc}        
\hline\hline                 
Sub-group & initial stellar mass &        age             & distance    & diameter & B stars\tablefootmark{a} & past SN\tablefootmark{b}\\
name         & $\sbr{M_\odot}$      & $\sbr{\rm Myr}$& $\sbr{\rm pc}$ & $\sbr{\rm pc}$ & $\sbr{\rm number}$      & $\sbr{\rm number}$\\
\hline
USco	         & 2060                        &	$\approx$~5-10\tablefootmark{d}	              &    145        &    35        &   49                               & 1\\
LCC		& $\approx 2000$\tablefootmark{c} & $\approx$~15                  &	    118        &    50          &   42                                & $\approx 4$\tablefootmark{e} \\
UCL		& $\approx 3000$\tablefootmark{c} & $\approx~$17                  &	    140            &    100          &   66                                & $\approx 7$ \\
\hline                                   
\end{tabular}
\tablefoot{ 
The data is taken from \citet{PreibMam08} unless indicated otherwise.
\tablefoottext{a}{Current number of B stars according to \citet{deBruijne99}.}
\tablefoottext{b}{Estimated number of past supernovae.}
\tablefoottext{c}{Estimated from the number of B stars with Salpeter initial mass function and scaled to the well determined mass of USco. 
For the initial mass function used in \citep{PreibMam08} for USco, USco would have lost only one B-star from the main sequence
while ageing from 5 to 16~Myr. }
\tablefoottext{d}{Stars with $M< 1$~\msun have age determinations towards the lower end of the range; more massive ones are 
found to be older \citep{HH15,PecMam16}. The range is likely related to measurement and model
uncertainties and not necessarily a real age spread \citep{Preibisch12,Donaldea17}.   }
\tablefoottext{e}{Scaled to the estimate in UCL using a Salpeter initial mass function.  }
}
\end{table*}


\begin{figure*}
  \centering
  \includegraphics[width=0.99\textwidth]{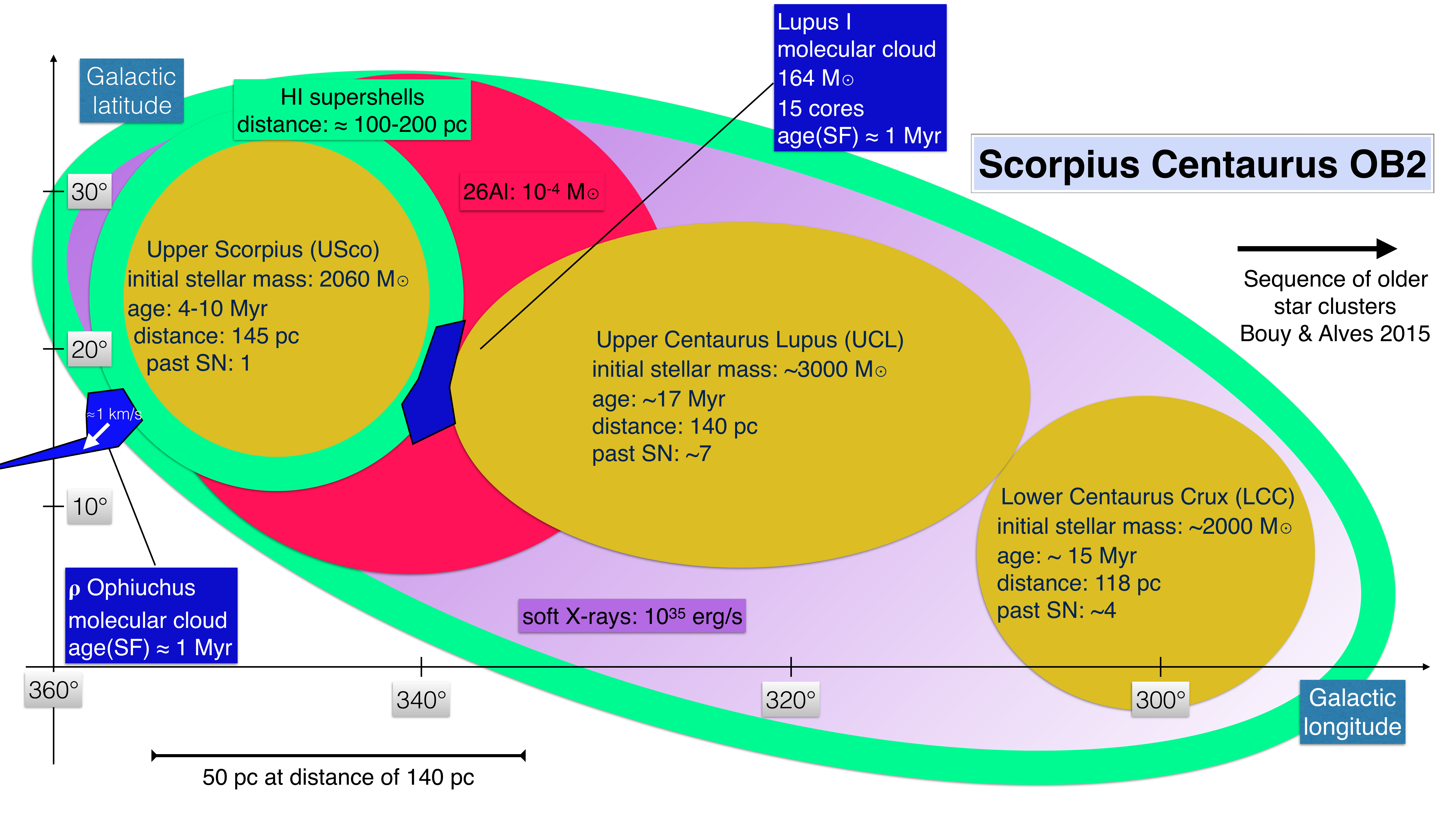}
  \caption{Summary sketch of observational information on the Scorpius-Centaurus region. 
	The OB-association Sco-Cen OB2 has three subgroups of $\approx 2000$~\msun, 
	each, formed over the last $\approx 15-20$~Myr. Stars are 
	currently forming in the $\rho$~Ophiuchus (mov\hi{ing} away from USco
	at about 1 \kms) and Lupus~I (part of an expanding
	\HI loop around USco) molecular clouds. See Table~\ref{t:sc-overview} for more details on the stars. 
	\HI shells are detected around the youngest OB subgroup, USco, and around the 
	entire region. Diffuse soft X-ray emission is detected towards the superbubble at a 
	level of $10^{35}$~erg~s$^{-1}$. One of the detected signatures of massive star winds 
	and supernova explosions is \al, towards USco.}

   \label{fig:scocensketch}%
\end{figure*}


\object{Scorpius-Centaurus OB2} with its associated superbubble (Sco-Cen in the following)
is the closest region 
with clear signs of recent massive-star activity \citep{PreibMam08}, 
and therefore an ideal object to study these 
effects in great detail. 
The OB association is at a distance of around 140~pc and 
consists of three subgroups, Upper Scorpius (USco), 
Upper Centaurus-Lupus (UCL), and Lower Centaurus-Crux (LCC) identified
from Hipparcos parallax and proper motion measurements \citep{deZeeuwea99,MelDam09},
\hi{which has recently been confirmed with \gaia data \citep{WM18}}.
The stellar populations are thus known in great detail \citep[e.g.,][]{deBruijne99,Preibea02,Mamea02}.
The three sub-groups currently contain many B~stars and, given their ages between 5 and 17~Myr, 
 probably had a number of more massive
stars in the past that exploded already 
\citep[e.g.,][Table~\ref{t:sc-overview} for basic stellar parameters]{deGeus92}. 

With an angular diameter of roughly $90\deg$ on the sky, Sco-Cen is also well studied at all
wavelengths that carry information about the diffuse interstellar medium (Fig.~\ref{fig:scocensketch}
for a schematic overview).
Large area and all-sky surveys, such as the \planck dust maps, or the ROSAT all-sky survey (details below)
show an essentially post star formation region. Most of the molecular gas has 
been converted to stars, or dislocated into an \HI supershell \citep{Poepea10}. 
The region is a prominent source at keV X-rays \citep{Gaczkea15}.
As X-ray bright superbubble, Sco-Cen likely hosted a supernova explosion
within the past Myr \citep{KD14,Krausea14a}. This is consistent with the detection
of radioactive \al towards one of the subgroups \citep{Diehlea10}.

Causal connection in the observed sequence of star
formation events has been suggested \citep[e.g.,][]{Preibea02}.  
\citet{PreibZinn07} detail a triggering scenario for the entire region:
Star formation first starts in UCL and LCC. Feedback in UCL produces an expanding 
high pressure region headed by a shock wave. The shock reached the USco parent cloud
about 10~Myr ago, pressurised the cloud and caused the denser parts of the cloud to collapse
and form stars. The cloud was subsequently eroded by its internal feedback which 
caused star formation in nearby clouds, e.g. the Lupus~I cloud and the \object{$\rho$ Ophiuchus} 
cloud. 

Several authors have suggested a possible link between the onset of star formation in the 
\object{$\rho$ Ophiuchus} region and the influence of expanding shells in Sco-Cen:
\citet{Wilkea79,Wilkea15} mention magnetic field structure and a higher velocity dispersion
of the YSOs in the main core of \object{$\rho$ Ophiuchus}, which could be a result from the 
global collapse occurring in the region due to ta shock wave from USco 
\citep[compare also][]{Kwonea15}. \citet{Klose86} argues for an enhanced gamma ray flux $>100$~MeV, which
he relates to an enhanced level of cosmic ray protons in the region, likely from an energetic explosion
which might have played a role in triggering star formation in \object{$\rho$ Ophiuchus}.
Proper motion studies show that USco has a projected velocity of
$(\mu_l \cos b, \mu_b)= (-24.5,-8.1)$~mas/yr \citep{deZeeuwea99}. Subtracting this from 
the projected velocity \hi{of} the \object{$\rho$ Ophiuchus} star cluster,
$(\mu_l \cos b, \mu_b)= (-23.8,-10.7)$~mas/yr \citep{Ducea17},
yields a relative motion of $(\mu_l \cos b, \mu_b)= (0.7,-2.6)$~mas/yr.
This translates to a projected velocity of 1.7~\kms roughly radially away from
USco, supporting the connection to feedback in USco.
%
\begin{figure*}
  \centering
  \includegraphics[angle=0,width=0.99\textwidth]{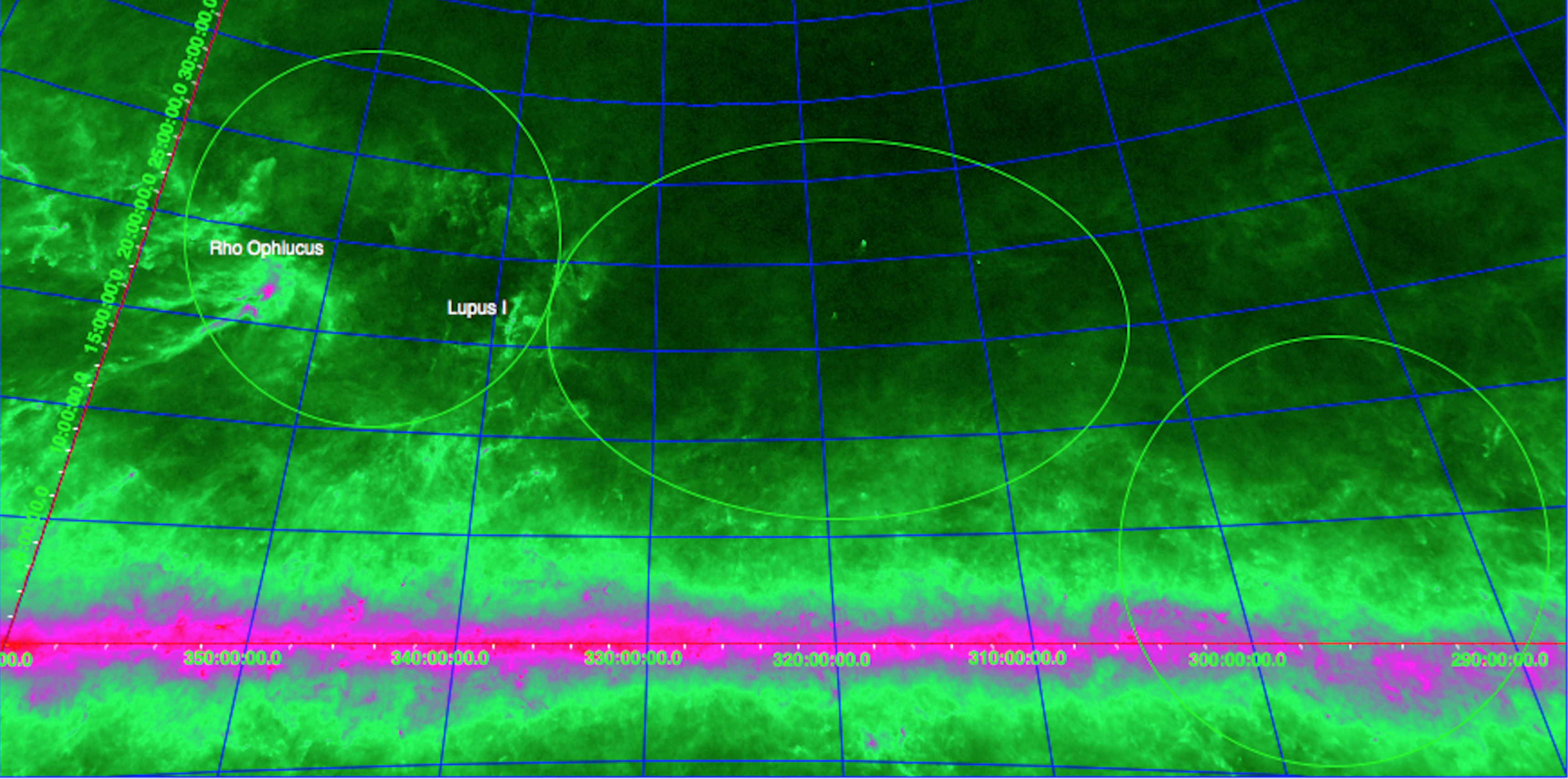}
  \caption{\planck 850$\mu$m dust map of the Scorpius-Centaurus region in Galactic coordinates.
	The Milky Way plane can be seen  as a pink band over the entire image.  
	Scorpius-Centaurus~OB2 (green ellipses indicate the location of the three subgroups)
	is mainly above the plane and essentially dust-free
	apart from the  $\rho$~Ophiuchus and Lupus~I clouds (labelled). Coordinates are Galactic and given in degree.
	The colour scale ranges from black (low) over green, pink, red, orange to yellow (high).}
   \label{fig:850mu}%
\end{figure*}

In a re-analysis of the {\sc Hipparcos} data, \citet{BouyAlves15} have recently expanded on this scenario
towards lower Galactic longitudes:
They find that the three Sco-Cen subgroups are part of a much larger stream of OB associations and 
open clusters which form a sequence of star formation over 60 Myr. Star formation would have started
65 Myr ago in the now open cluster \object{NGC 2451A} and propagated 350~pc along a filament
with the last major episode in USco. Discrete star formation events took place every
few to 30 Myr.

While the basic case for sequential star formation in Sco-Cen is strong, 
details remain hard to understand. For example, if each star formation event sweeps up
the dense interstellar medium in its surroundings before it triggers the formation
of another OB association in its shell \citep[collect \& collapse, e.g.,][]{Whitwea94}, one 
would expect a monotonic age increase towards the oldest stars, which is not seen in Sco-Cen.
This is similar for a distribution of clouds close to equilibrium between self-gravity and pressure
that might be triggered to collapse by a passing shock wave.
If such a distribution of clouds existed for a time similar to the age of the Sco-Cen stars,
one would expect at least some of them to form stars independently of any trigger.

Here we compile multiwavelength data on the interstellar medium in Sco-Cen,
and interpret these with the help of dedicated superbubble simulation\hi{s}.
We find that simple collect and collapse scenarios are disfavoured. Even the smallest,
currently star-forming molecular cloud, shows evidence for a multi-stage formation process 
where cloud formation and the onset of star formation are independent from each other.
Our simulations show that an initially homogeneous medium cannot fully explain the
data. We suggest a multi-stage formation process where the high sound speed
in a hot superbubble plays an important role in communicating pressure enhancements
to different parts of an initially flattened linear cloud inside the superbubble.

\begin{figure*}
  \centering
  \includegraphics[width=\textwidth]{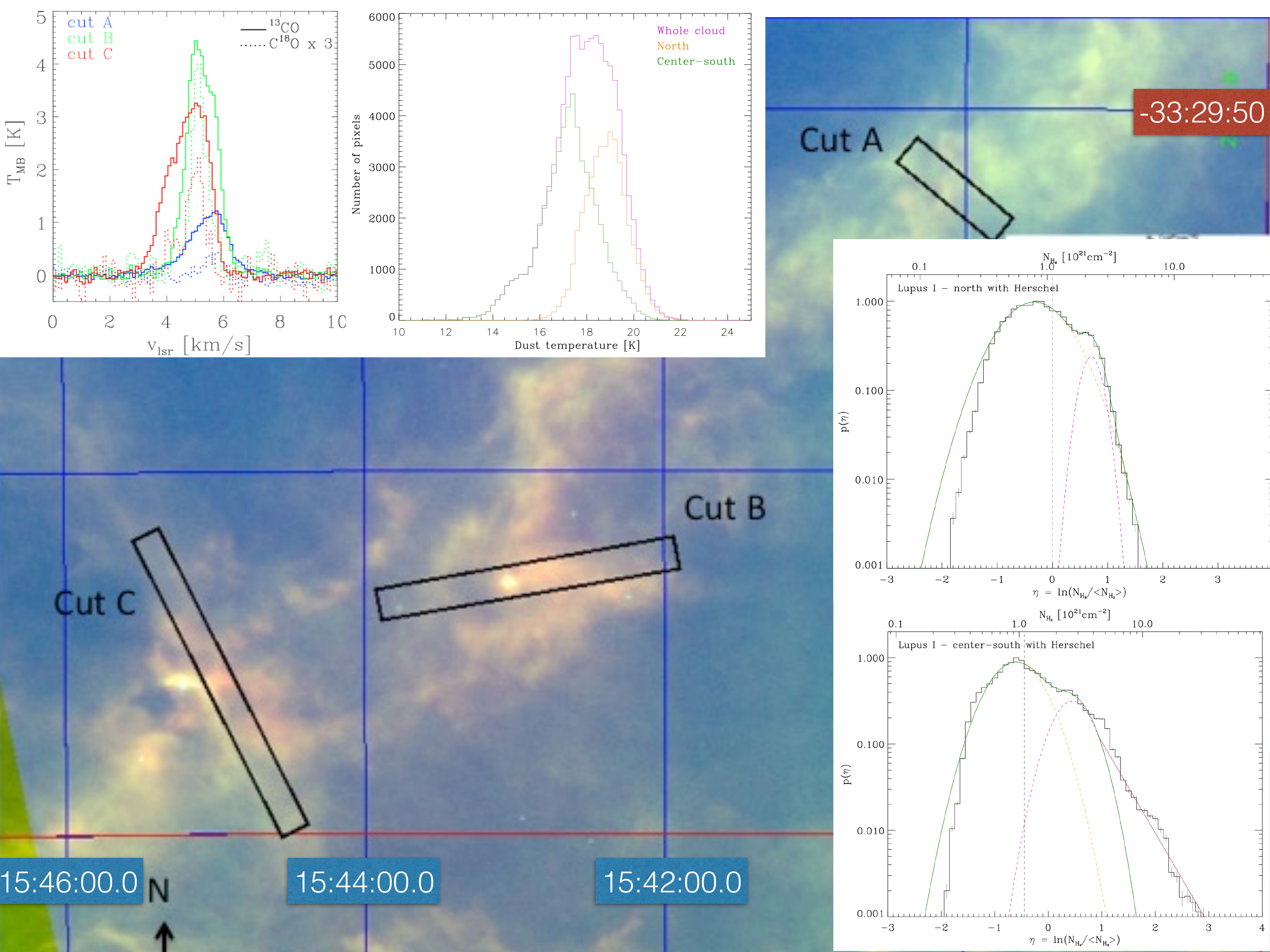}
    \caption{Lupus I molecular cloud, 
	combined image of the mosaics obtained with {\it SPITZER}/MIPS at 
	24~$\mu$m (blue), \heru/PACS at 160~$\mu$m (green) and the 
	\heru/SPIRE at 500~$\mu$m (red), all in logarithmic scales. 
	The right overplotted panels show the column density distribution functions 
	of the northern (top) and southern (bottom) parts of LupusI and the central 
	panel shows the histograms of the dust temperature in different parts of the 
	cloud \citep[see details in][]{Gaczkea15}. The left panel shows the gas velocity 
	profiles averaged along cuts A, B and C, which are highlighted by the black boxes.	
	Ecliptic coordinates are used for this figure to help identifying northern and 
	southern parts. Galactic coordinates of the centre of Lupus~I: $l=338\deg 50\arcmin$,
	$b=16\deg 40\arcmin$.
	}
   \label{fig:lupIapex}%
\end{figure*}

\begin{figure*}
  \centering
 \includegraphics[width=0.49\textwidth]{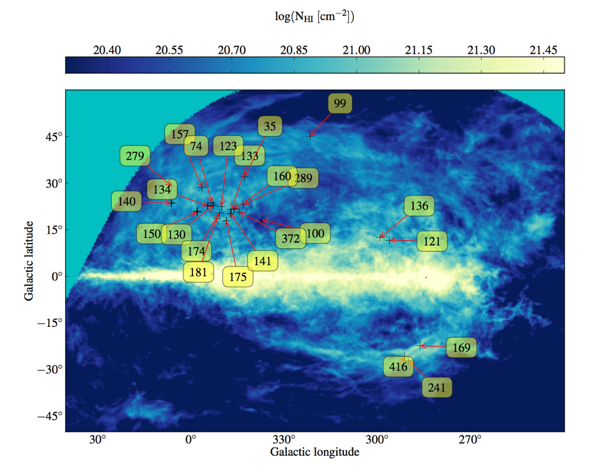}
  \includegraphics[width=0.49\textwidth]{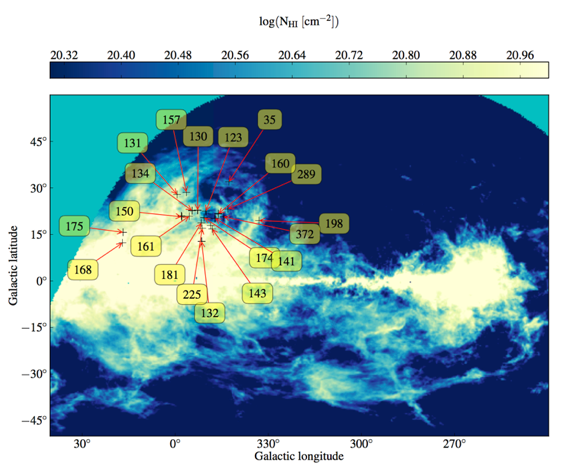}
    \caption{Intensity of the 21cm line of neutral hydrogen towards Sco-Cen.
	The left (right) panel shows the emission integrated between -20~\kms and 0 (0 and 10~\kms).
	The numbers in the yellow boxes show upper distance limits for the \HI features 
	derived from velocity-matched \NaI absorption lines against stars with distances
	known from parallax measurements by {\em Hipparcos} and {\em GAIA}.
	}
   \label{fig:HI-big}%
\end{figure*}

\begin{figure*}
  \centering
 \includegraphics[width=0.4\textwidth]{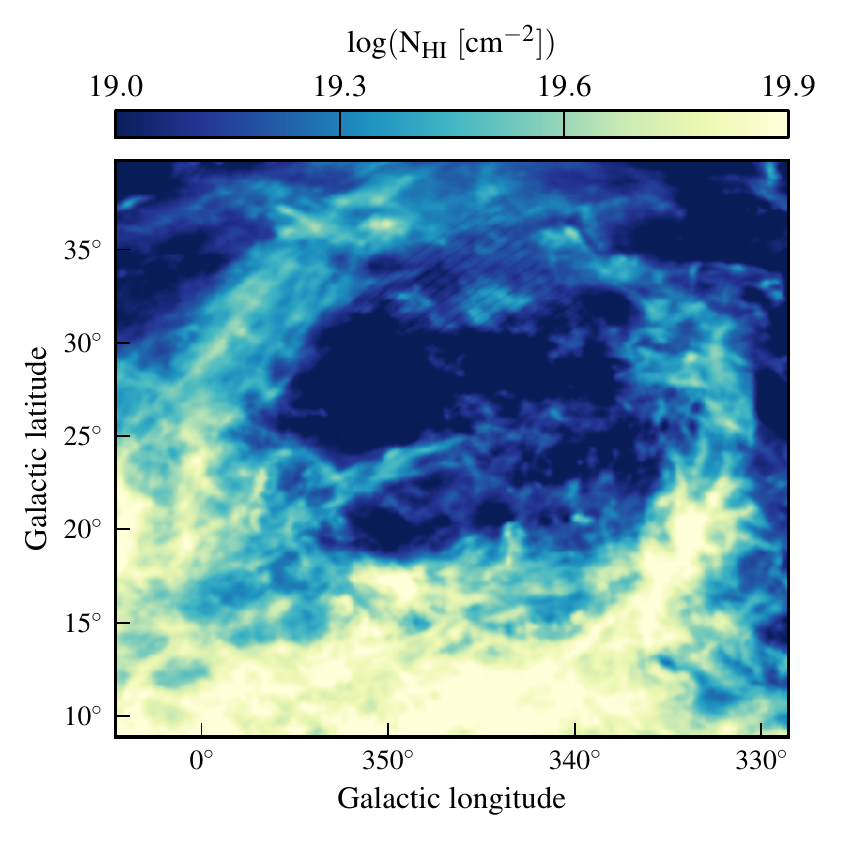}
  \includegraphics[width=0.5\textwidth]{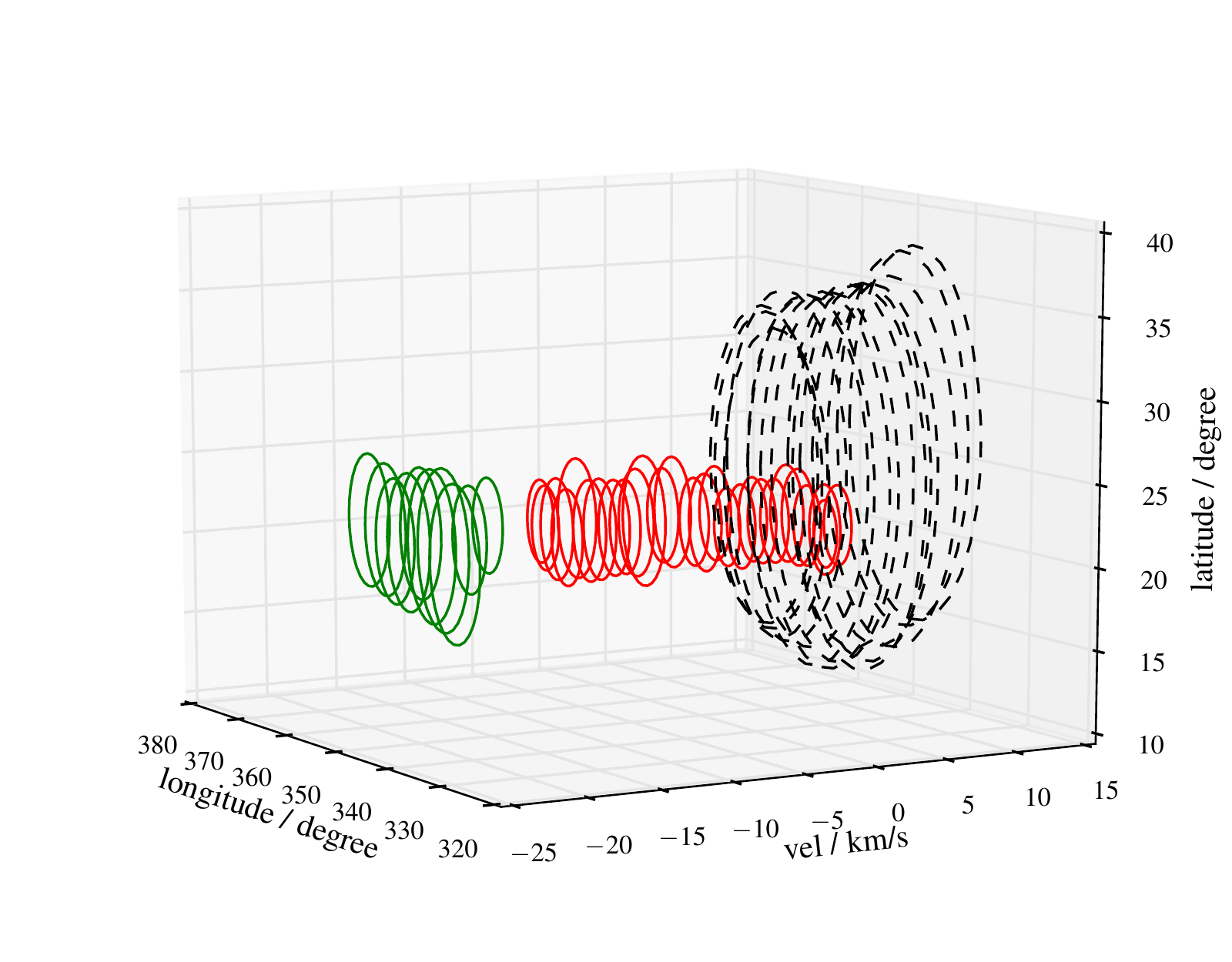}
    \caption{Intensity of the 21cm line of neutral hydrogen towards USco in the 7.4~km~s$^{-1}$
	velocity channel (left) and by-eye characterisation
	of the \HI structures across multiple velocity channels (right). 
	Three groups of rings are apparent:
	the USco loop (black), which corresponds to the dominant loop in the 
	channel map on the left,
	and two smaller, tube-like structures (green and red) which are cospatial 
	but separated in velocity space. The red and green rings might 
	be a result of erosion of the \HI interface between the Sco-Cen superbubble 
	and the local bubble, where the hot gas streaming through the hole
	ablates and accelerates the \HI. 
	}
   \label{fig:HI-USco}%
\end{figure*}
\begin{figure}
  \centering
 \includegraphics[width=0.5\textwidth]{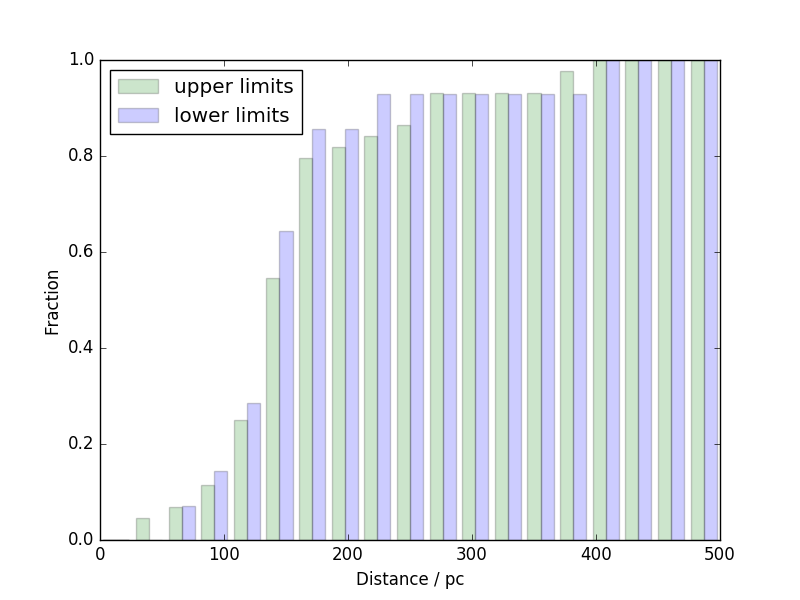}
  \includegraphics[width=0.5\textwidth]{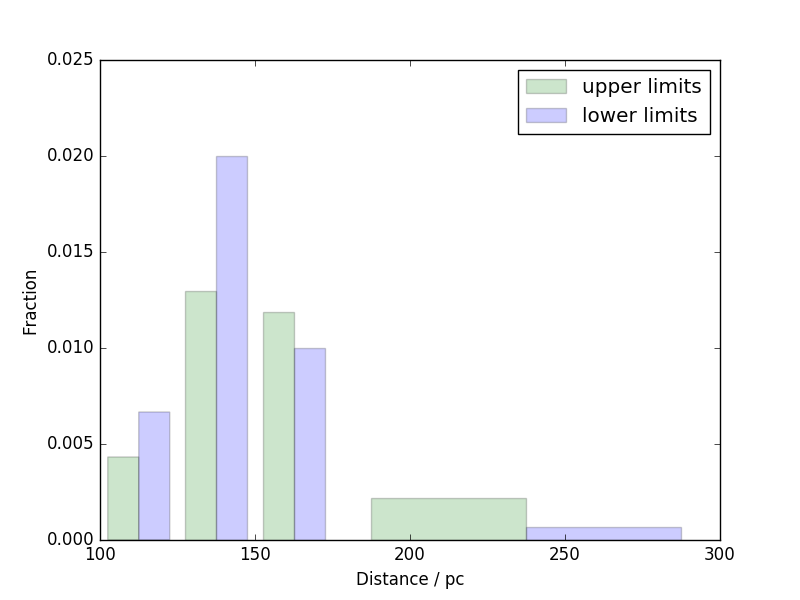}
    \caption{Histograms of distances of HI features in Sco-Cen determined by 
	association with \NaI absorption lines. Top: cumulative histogram for all
	available stars. Bottom histogram \hi{from} stars at distances between 100 and 300~pc.
	More lower (upper) limits are found up to (above) about 150~pc, 
	Almost no lower limits are found above 220~pc.
	}
   \label{fig:HI-histos}%
\end{figure}

\section{Observations and Simulations}\label{sec:o_and_s}

We summarise observations of dust and 
molecular gas which have already been discussed in
\citet{Gaczkea15,Gaczkea17}, and present new analyses for  
atomic hydrogen, hot X-ray gas and gamma rays from
the radioactive trace element
\alu . We also performed new 3D hydrodynamics simulation\hi{s}
specifically for the Sco-Cen superbubble. 

\subsection{Molecules \& dust}\label{sec:mol}
 We summarise here our far-infrared and sub-millimeter results for the 
entire Sco-Cen region including the Lupus I molecular cloud 
which is placed between the USco and UCL subgroups.
Details can be found in \citet{Gaczkea15,Gaczkea17}.

In order to trace the {\em large-scale dust} distribution of the cold dust of the 
Sco-Cen region we used the archival  850~$\mu$m  dust map provided by the 
\planck mission (High Frequency Instrument, \planck legacy archive release 
PR1 21.03.2013) with a resolution of 5\arcmin (Fig.~\ref{fig:850mu}).

The {\em small scale dust} observations of the Lupus I molecular cloud were collected in the 
far-infrared using archival data from the \her at 70, 160, 250, 350, 
and 500~$\mu$m. The angular resolution was 3.2\arcsec , 4.5\arcsec , 6\arcsec , 
8\arcsec , and 11.5\arcsec 
in the five maps, respectively \citep[details in][]{Gaczkea15}.
We obtained 870~$\mu$m maps with the LABOCA bolometer  at the APEX 12~m 
telescope with an angular resolution is 19.2\arcsec , and the total field of view 
is 11.4\arcmin \citep[][and references therein]{Gaczkea15}.
Observations along three different scans through the Lupus I cloud  simultaneously in the 
$^{13}$CO(2-1) and C$^{18}$O(2-1) lines using the APEX-1 receiver obtained a  
final angular resolution of 30.1\arcsec,
which corresponds to 0.02~pc at the distance of Lupus~I of 140~pc \citep{Gacz16,Gaczkea17}.
Observational results on the Lupus~I cloud are summarised in Fig.~\ref{fig:lupIapex}.

Figure~\ref{fig:850mu} shows the 850~$\mu$m \planck dust map. The OB-association 
is located mainly above the galactic plane and essentially free of dust, as it is expected for 
a star-forming region after a few Myr. The cold molecular gas left is concentrated in the Lupus I and 
$\rho$~Ophiuchus clouds and few filamentary structures.
Near $\rho$~Ophiuchus, filaments extend preferentially towards decreasing 
Galactic longitudes and latitudes. This suggests a large-scale stream of 
tenuous gas towards the lower left that ablates clouds in this 
direction, leading to their elongated appearance. 
Other filaments might be parts of shells of swept-up interstellar
medium.

The Lupus I molecular cloud (Fig.~\ref{fig:lupIapex}) is spatially and kinematically associated with an 
\HI shell around USco, and is located, in projection, on the connecting line between USco and UCL. 

The northern part of Lupus I has lower densities, higher temperatures, and no 
active star formation, while the center-south part harbours dozens of pre-stellar 
cores where density and temperature reach their maximum and minimum, respectively. 
The analysis of the column density probability distribution functions (PDFs) from 
\her data show double-peaked profiles for all parts of the cloud. We attribute this
to an external compression event, after the lognormal PDF had first been established,
quite possibly by turbulence associated with thin-shell instabilities.  
Increase of ambient pressure as expected for example from a supernova in USco
could plausibly have compressed part of the Lupus~I cloud and shifted its PDF to higher densities.  
In those parts with active star formation, the PDFs show 
a power-law tail.
The PDFs we calculated from our LABOCA data trace the denser parts of the cloud 
showing one peak and a power-law tail. With LABOCA we identified 15 cores with 
masses between 0.07 and 1.71~\msun~and a total mass of $\approx 8$~\msun. 
The mass of the cores represents $\approx5$\% of the total gas and dust mass of 
the cloud of $\approx 172$~\msun~(average between the \her and \planck 
resulting dust mass). 

$^{13}$CO(2-1) and C$^{18}$O(2-1) line observations of Lupus I with the 
APEX telescope at three distinct scans through the cloud show a complex 
kinematic structure with several line-of-sight components that may overlay each other. 
Such complex kinematic structure is expected for turbulence.
LTE analysis showed that the C$^{18}$O line is optically thin almost 
everywhere within the three scanned regions, in contrast to the $^{13}$CO line. 
We therefore used the 
C$^{18}$O measurements to assess the kinematics. 
The non-thermal velocity dispersion is in the transonic regime in all parts of 
the cloud. This level of turbulence is expected from instabilities on a decelerating shell
\citep{Krausea13a}.

\subsection{Atomic gas}
We analysed the entire Sco-Cen region in \HI using the Parkes Galactic all-sky survey
\citep[GASS,][]{McClurea09,Kalbea10}\footnote{https://www.astro.uni- bonn.de/hisurvey/gass/}. 
GASS has a spatial resolution of 16.2\arcmin  \citep{KalHau15}, 
which corresponds to a length scale of 0.66 pc at the 
characteristic distance of Lupus I. The velocity resolution is 1~\kms. 
We focused in particular on the prominent shell around USco
where the GASS observations reveal finer detail than previous observations \citep[compare][]{Poepea10}.

In order to assess the distances of \HI features, we compared their characteristic velocities with Na-D
absorption features measured against stars with known distances.
We use catalogued absorption features from \citet{Poepea10} and 
\citet{Welshea10}\footnote{This study included a much higher number of sightlines, 
but due to a dataloss the velocity values of the peaks were not available (priv. com. with B.Y. Welsh 2014).}.
Distances were calculated from parallax measurements with 
{\em HIPPARCOS} with distance uncertainties of $\approx 7$ per cent for 50-100~pc, 
and $\approx 15$ per cent for 200-250~pc distances.
Where available, more accurate \gaia parallax measurements were used
\footnote{The data were downloaded using the \gaia web archive: https://gea.esac.esa.int/archive/ at} 
\citep{GAIA16a,Lindegrea16}. In the present data release~1, parallax uncertainties are typically 0.3~mas (0.3~mas) 
for statistical (systematical) uncertainties, which translates to a statistical error of 5~per cent at a distance of 150~pc.
For most cases the distance estimation 
using parallax measurements led to similar results but there are also some cases where the HIPPARCOS 
and the \gaia distances show significant differences, e.g. the star HD 146284 which has a distance of 
264 pc according to the HIPPARCOS data and one of just one of 181.23 pc according to the \gaia data. 
\gaia distances were adopted where available. 

We associated \NaI lines with \HI features based on the \HI profile and 
cap linewidths at 3~\kms~to ensure that the line peak is not 
fitted to a broad \HI component that reflects the galactic disc background. If the 
\NaI absorption line fits more than one \HI component, the closer match is selected.
 
In this way, we derive upper limits for the distances to \HI features, where a star shows
associated absorption, and lower limits where a star shows no associated \NaI absorption feature.

We integrate the \HI data cubes over suitably selected velocity channels and
show these integrated channel maps in Fig.~\ref{fig:HI-big}. They show shells and arcs
as reported already by \citet{Poepea10}, as well as a high intensity in the Galactic plane.
We discern two main features of interest: Firstly, a big supershell which surrounds the entire region. It is located
in an oblique way with respect to the Galactic plane, at greater latitudes for greater longitudes.
This morphology is similar to the one of the Sco-Cen subgroups and encloses them in projection.
It is expected that superbubbles within the disc scale height 
\citep[$\approx 150$~pc for HI,][]{NJ02}
are oriented 
randomly with respect to the Galactic disc plane.

The second prominent structure is the USco loop, which we show in more detail in a single-channel map 
Fig.~\ref{fig:HI-USco}.
The loop is plausibly originating from the USco stars. We show its 3D structure with velocity as the third dimension
also in Fig.~\ref{fig:HI-USco}. Circles were fitted by eye to the individual channel maps.
The loop appears to have roughly constant diameter in all velocity channels, suggesting that it
has experienced blow-out into (or away from) the line of sight, and only its cylindrical central parts remain visible.

We have highlighted another loop-like structure in the 3D representation in Fig.~\ref{fig:HI-USco}. This smaller
loop is seen towards approaching velocities and towards USco. It could be a signature of erosion of the interstellar 
medium between the Sco-Cen superbubble and the local superbubble.

For the distances to the \HI features in the Sco-Cen region, 
we have obtained 14 lower limits and 44 upper limits by associating the \HI features 
to \NaI absorption lines.
We show, for clarity, only the upper limits in Fig.~\ref{fig:HI-big}. From a statistical analysis
(Fig.~\ref{fig:HI-histos}), we find \HI features at a range of distances, bounded by 
100~pc from below and about 250~pc above. The more detailed histogram for distances between 100 and 300 pc
shows that we find more lower limits below 150~pc, and more upper limits for above 150~pc. This suggests that
the \HI features are located at a range of distances centred around a characteristic value of 150~pc. 
Because of the distance limit in the \NaI database, we can only say
that the supershell extends to at least 220~pc. This statistical analysis is consistent with the distances
for individual features determined by \citet{Poepea10}.

Using these distances, we can determine approximate sizes of the \HI features. The big supershell
surrounding the entire Sco-Cen region has about $90^\circ$ diameter, corresponding to 240~pc at
a distance of 150~pc. The USco loop has $25^\circ$ diameter, which corresponds to 65~pc.
The \HI shell itself is spatially resolved and has a thickness of about $4\deg$, corresponding to about
10~pc. It has a filamentary appearance. Thick filamentary shells are expected from 3D simulations
of superbubbles that resolve the clumping instabilities in the shell \citep{Krausea13a}.
\rvb{The sonic random motions observed in the CO lines in part of the shell (Lupus~I, compare Sect.~\ref{sec:mol})
would also be expected as a result of the instability. Detailed estimates of unstable wavelengths and timescales
are difficult for this case, as the blowout has likely modified the deceleration history of the shell, which is a 
crucial parameter \citep[e.g.,][]{VR89}. Please, see Sect.~\ref{sec:itp} for further discussion.}

\subsection{Hot gas}\label{s:xray}

\begin{figure}
  \centering
  \rotatebox{0}{\includegraphics[width=0.49\textwidth]{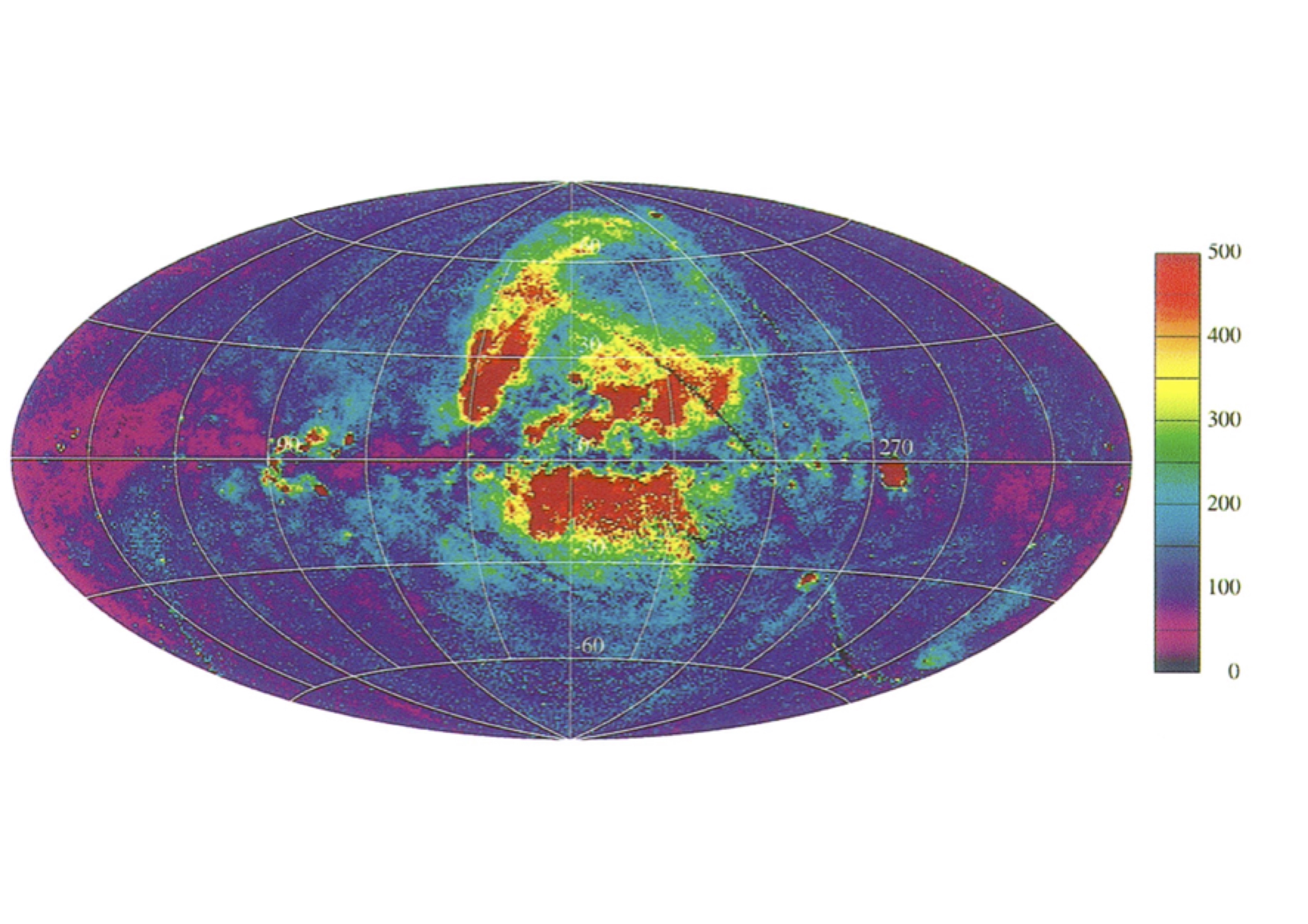}}
  \caption{3/4 keV map from the \rosat all sky survey \citep{Snowdea97}. 
	Prominent emission is seen at $300\deg<l<360\deg$, $0<b<30\deg$, spatially coincident
	with the major part of the Sco-Cen region. The emission below the Galactic plane
	might be related. 
	}
   \label{fig:X}%
\end{figure}

The ROSAT all sky survey \citep[Max-Planck-Institut f\"ur extraterrestrische Physik (MPE) archive]{Snowdea97} 
shows X-ray emission from hot plasma in the interiors of the above cavities.
We determined the count rate in the area of interest in ROSAT 
bands 4-7 from the maps and converted it to an equivalent energy flux, assuming thermal 
bremsstrahlung of diffuse gas. We used 
a typical value for the hot gas temperature in superbubbles \citep{Krausea14a}, 0.5~keV.
The conversion factor  from diffuse X-ray flux to gas temperature was obtained from 
Fig.~10.9 of the ROSAT technical appendix, available from MPE.

The Sco-Cen superbubble is clearly detected in X-rays (Fig.~\ref{fig:X}).
It is difficult to make an accurate quantitative analysis of the bubble, 
because it is located close to the 
Galactic plane and hence possibly affected by absorption and source confusion, i.e.,
foreground and background emission. The detection in the 3/4~keV band
suggests a sub-keV temperature. With the procedure described above, we obtain an indicative 
luminosity of $\approx 2\times10^{35}$~erg~s$^{-1}$ for the region $300\deg<l<360\deg$, $0<b<30\deg$. This is
towards the lower end of the range occupied by X-ray-bright superbubbles,
and hints at a recent ($\approx 1$~Myr) supernova \citep[compare][]{Krausea14a}.


\begin{figure}
  \centering
  \includegraphics[angle=-90,width=0.49\textwidth]{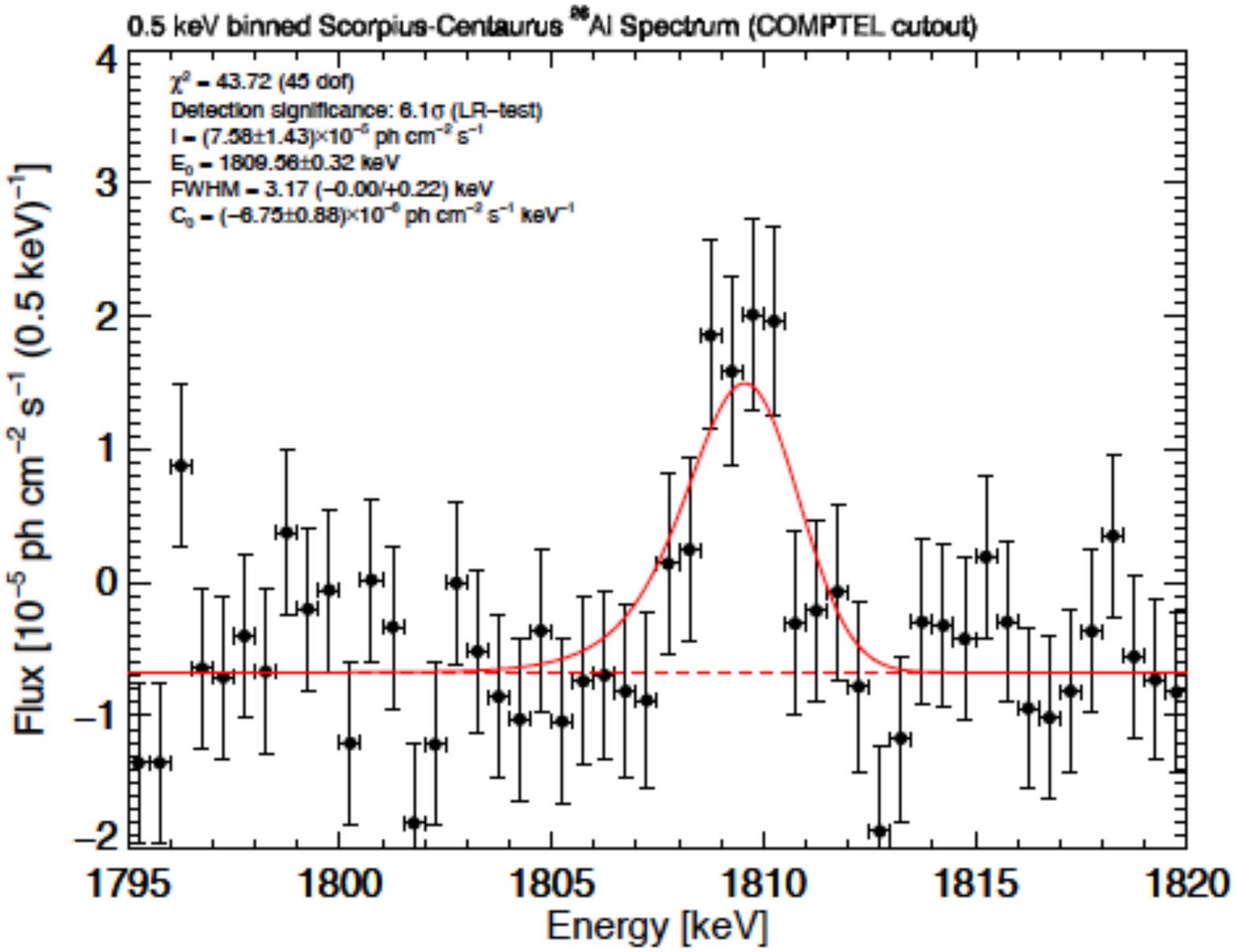}
  \caption{10Msec-INTEGRAL spectrum towards Scorpius-Centaurus in energy bands covering
	the 1809~keV radioactive decay line of \al. The signal corresponds to a $6\sigma$ detection,
	and is consistent with the nucleosynthesis yields of several massive stars.
	}
   \label{fig:26Al}%
\end{figure}

\newlength{\flength}
\setlength{\flength}{0.246\textwidth}
\begin{figure*}
  \centering
 \includegraphics[angle=-0,width=\flength]{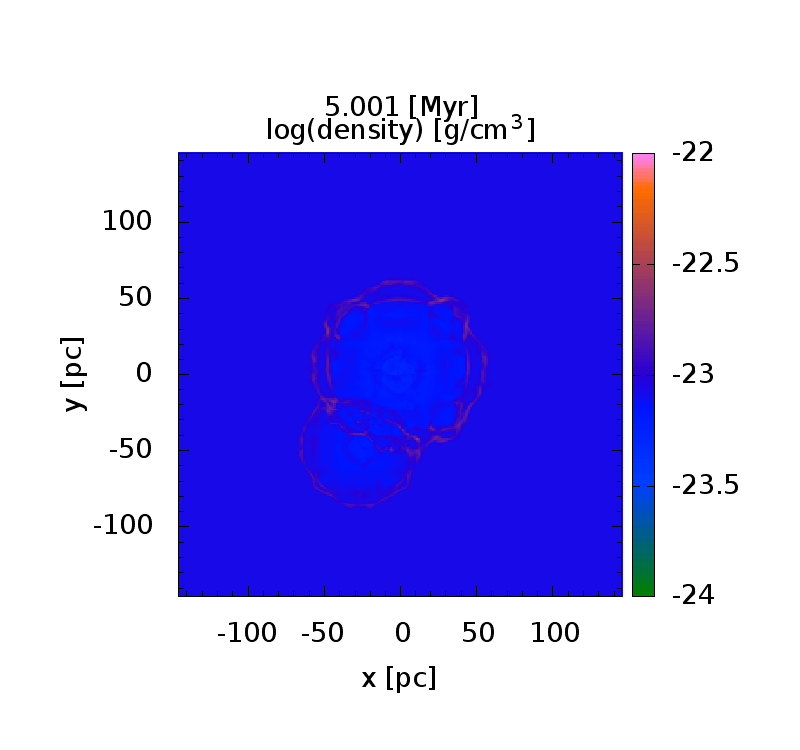}
  \includegraphics[angle=-0,width=\flength]{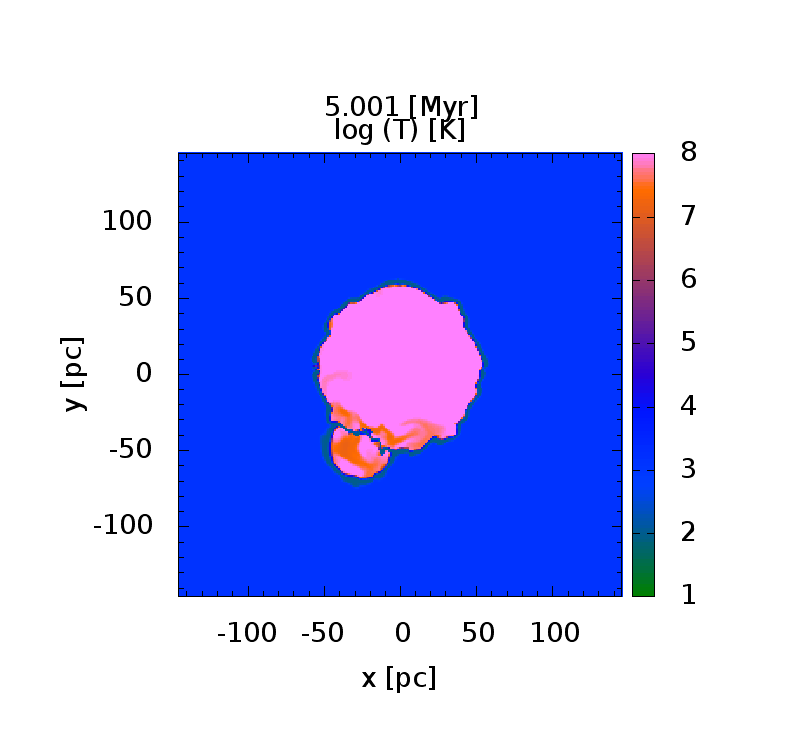}
  \includegraphics[angle=-0,width=\flength]{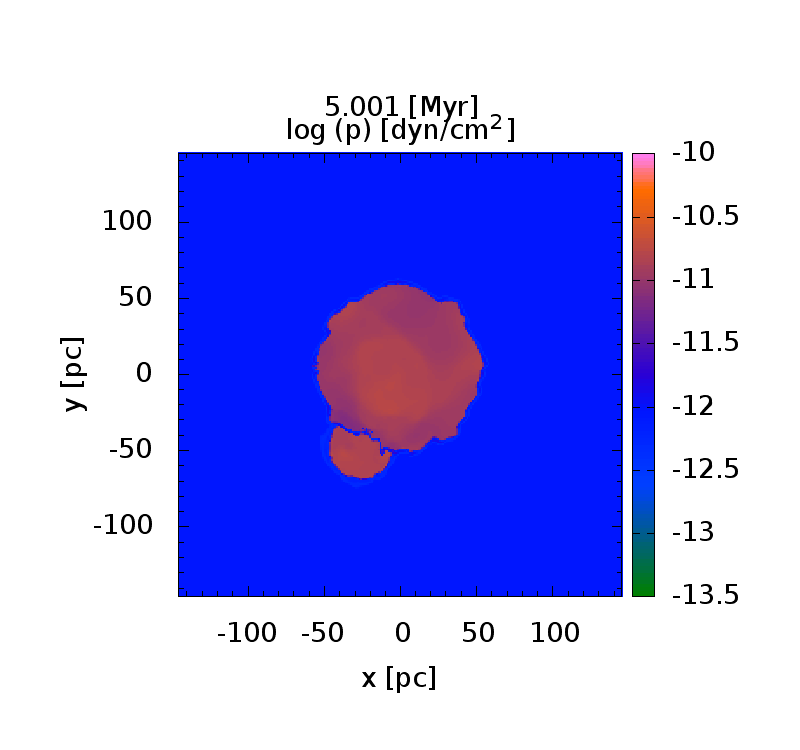}
  \includegraphics[angle=-0,width=\flength]{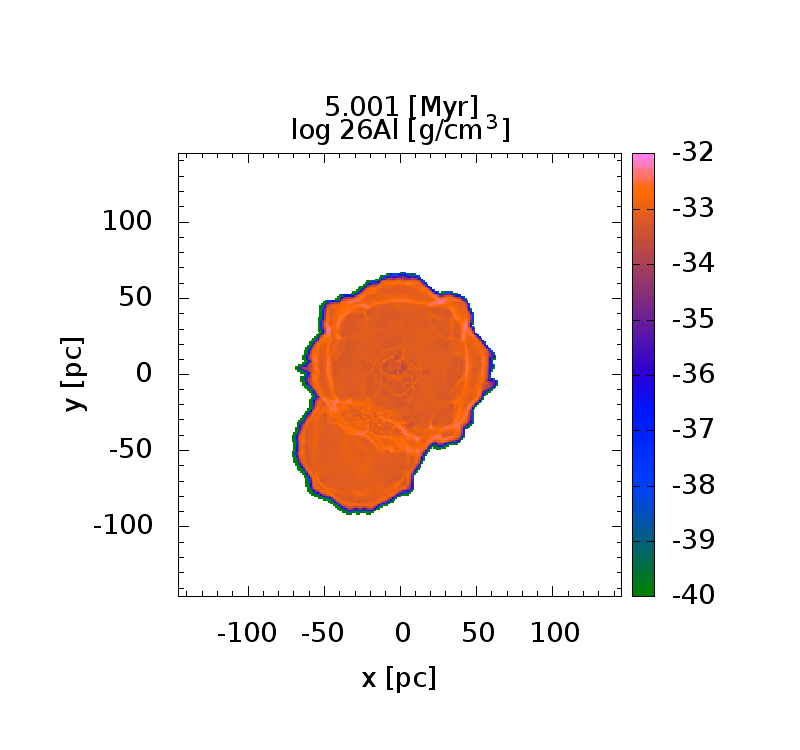}
  \includegraphics[angle=-0,width=\flength]{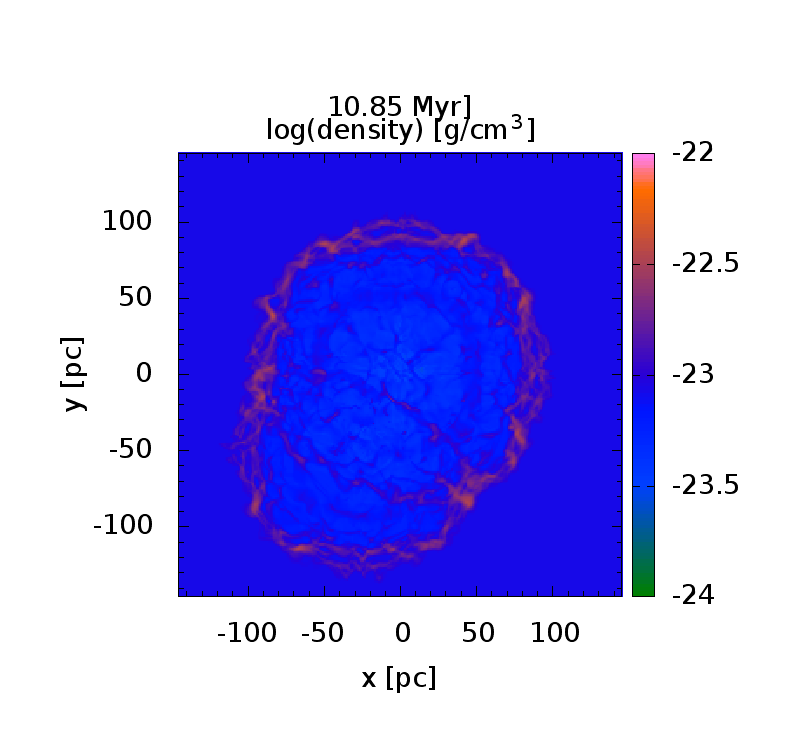}
  \includegraphics[angle=-0,width=\flength]{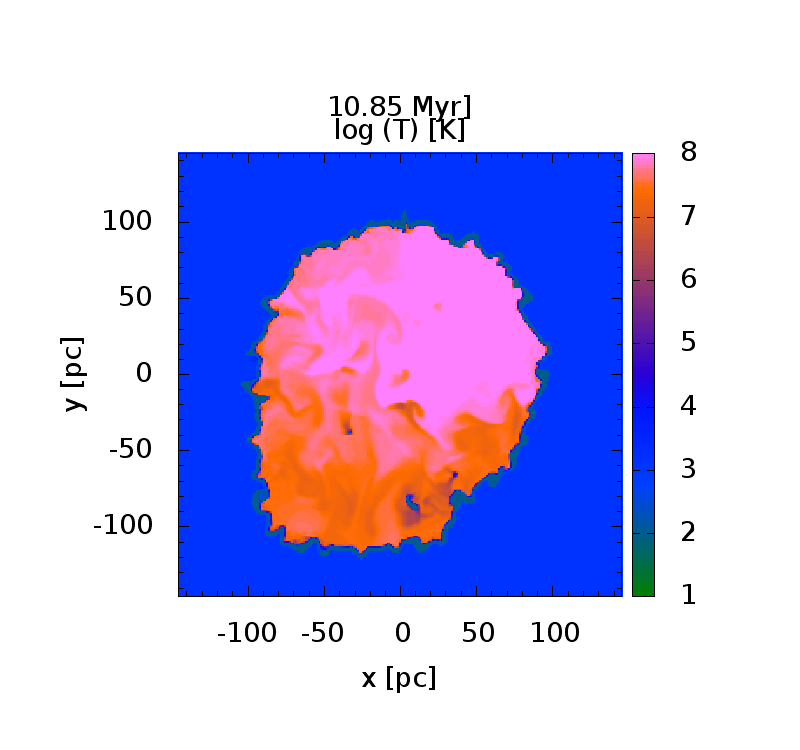}
 \includegraphics[angle=-0,width=\flength]{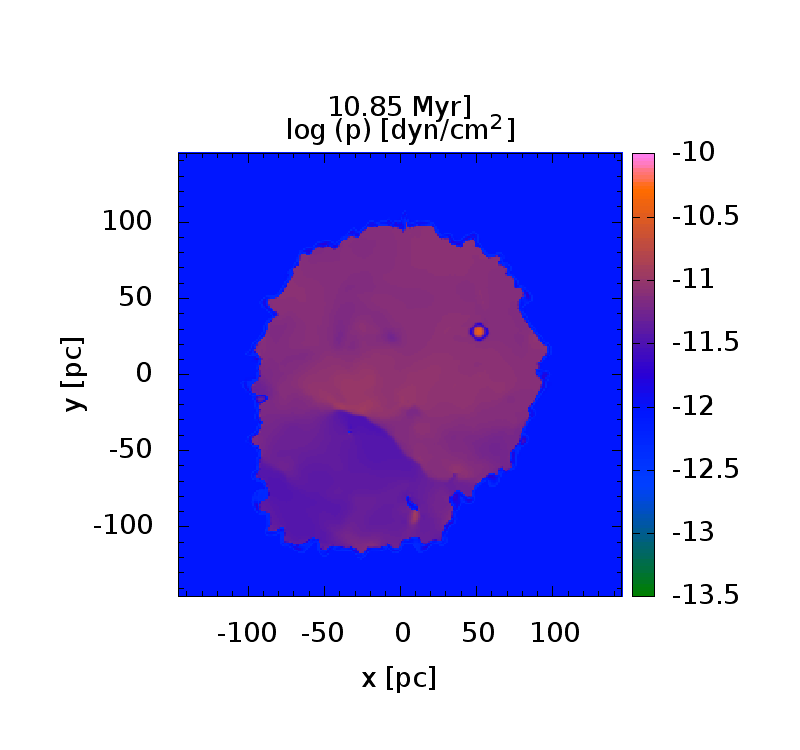}
  \includegraphics[angle=-0,width=\flength]{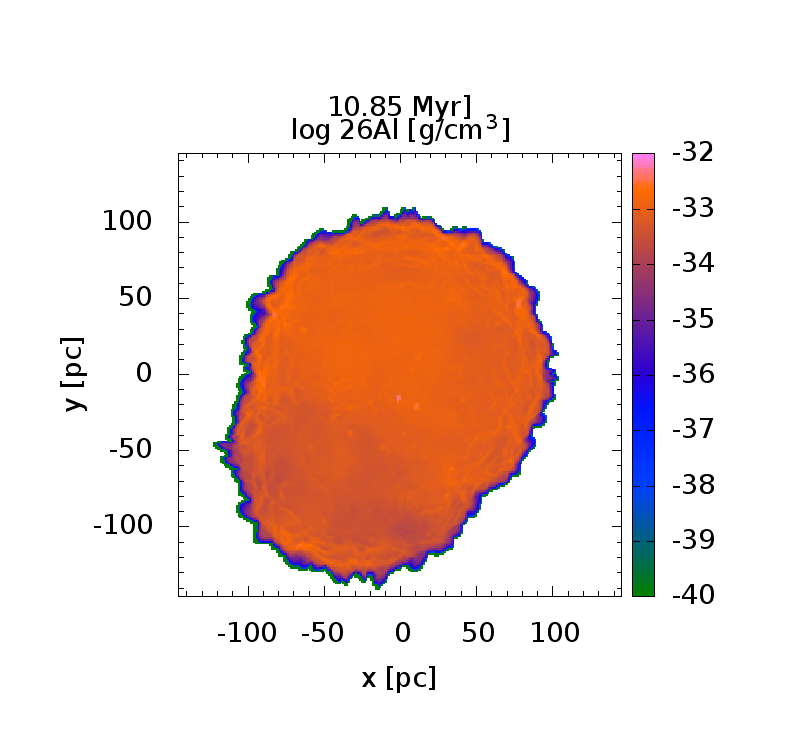}
  \includegraphics[angle=-0,width=\flength]{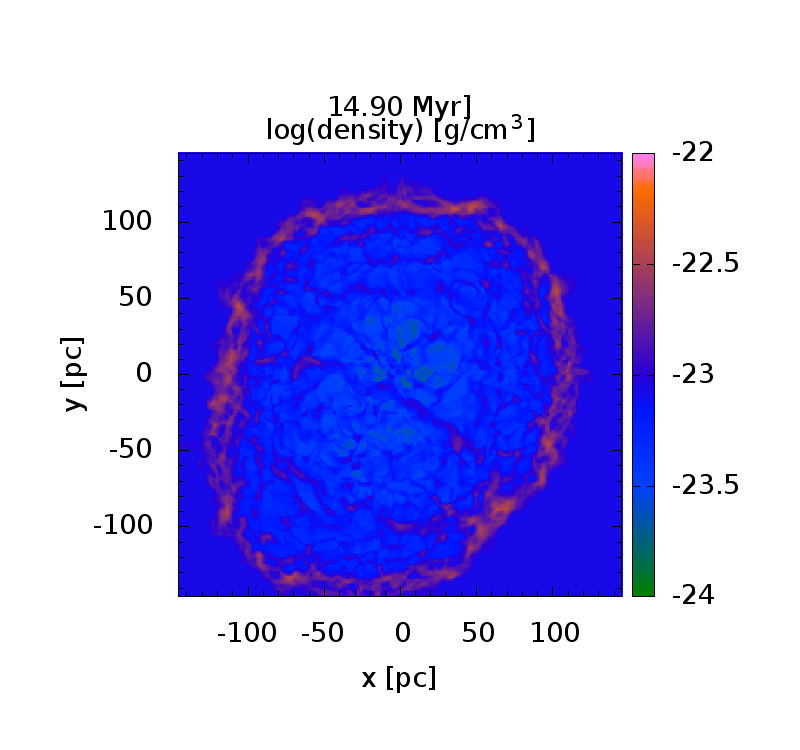}
  \includegraphics[angle=-0,width=\flength]{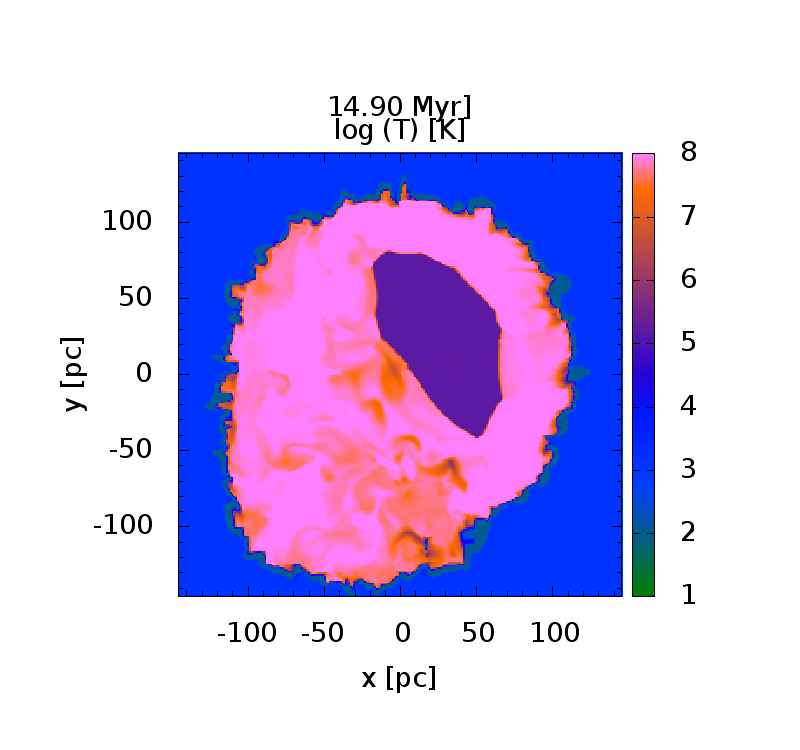}
  \includegraphics[angle=-0,width=\flength]{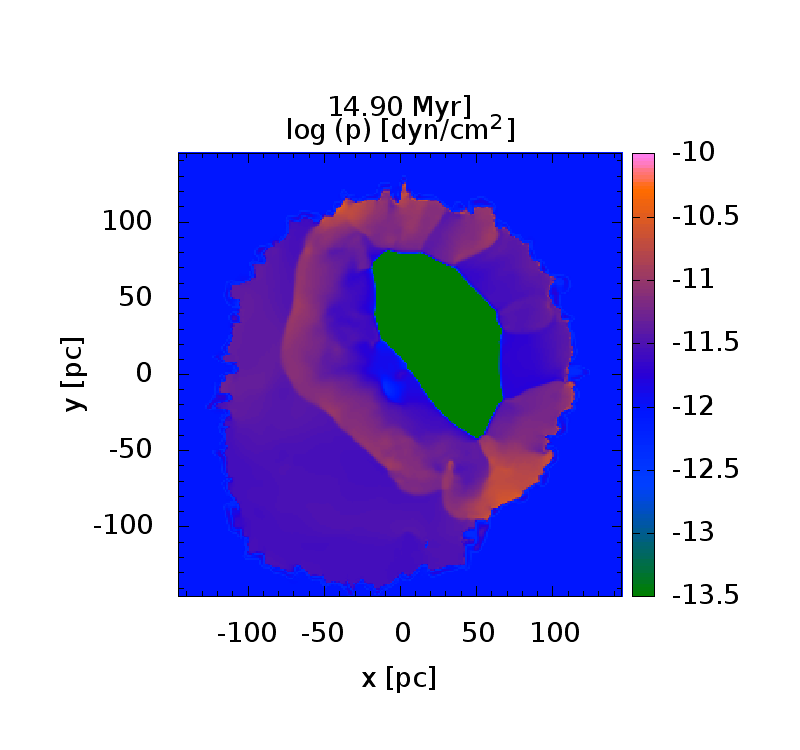}
  \includegraphics[angle=-0,width=\flength]{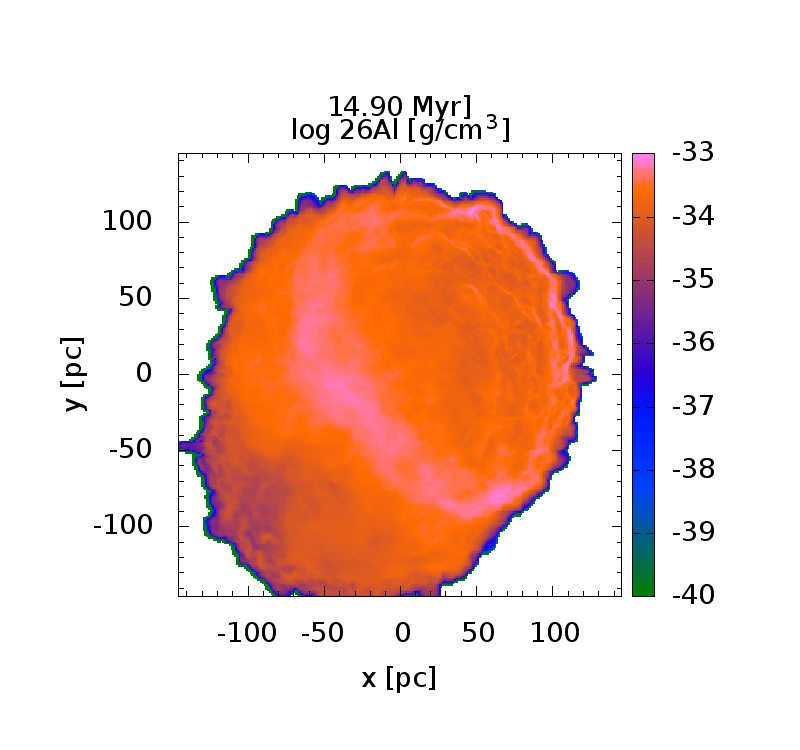}
  \includegraphics[angle=-0,width=\flength]{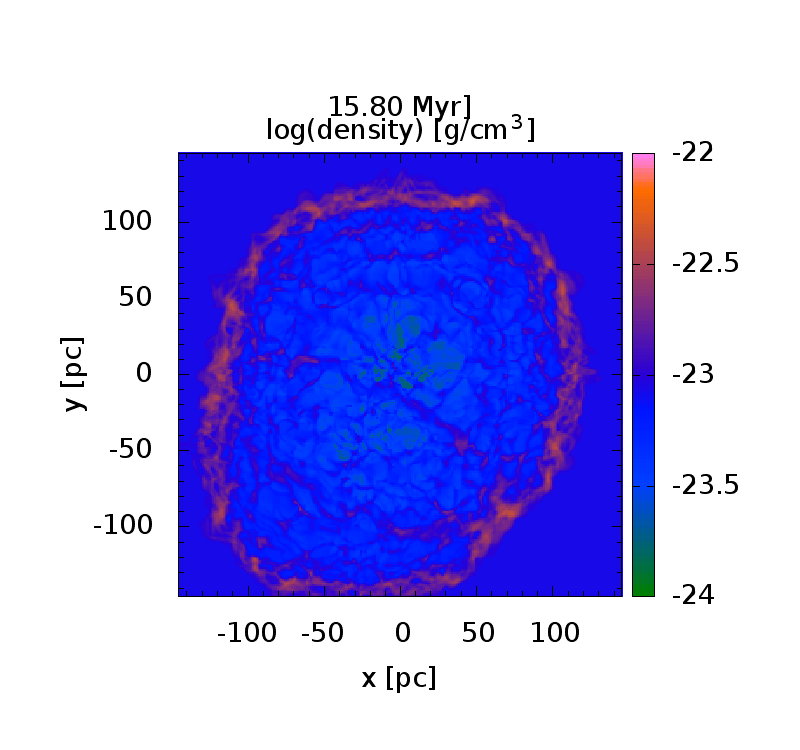}
  \includegraphics[angle=-0,width=\flength]{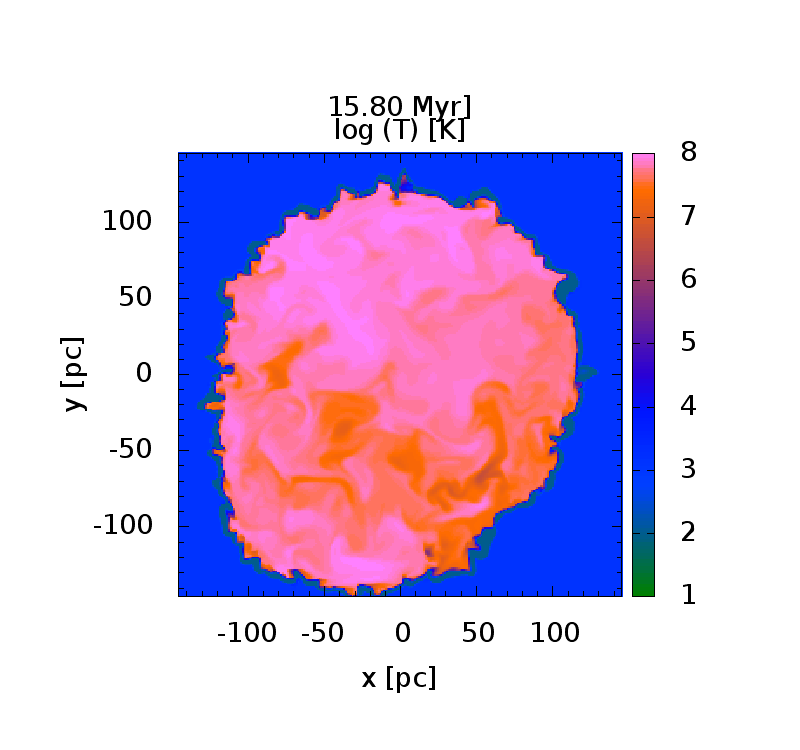}
  \includegraphics[angle=-0,width=\flength]{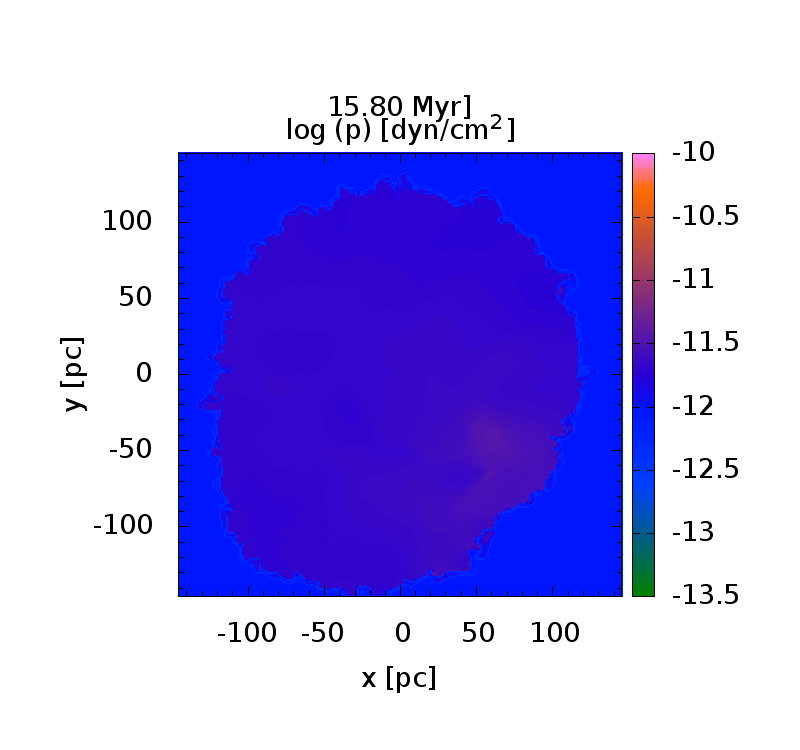}
  \includegraphics[angle=-0,width=\flength]{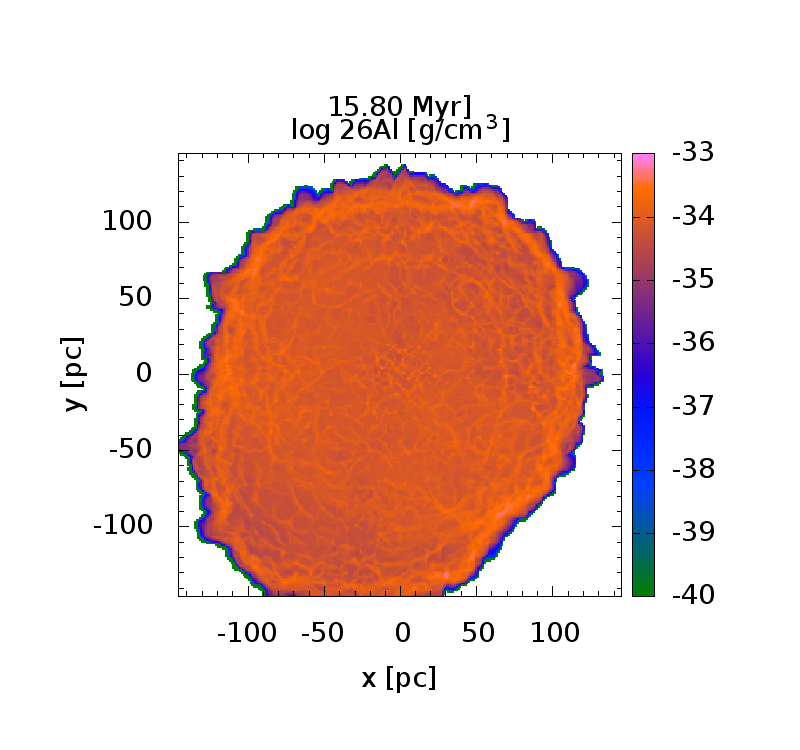}
\caption{Time evolution of the 3D hydrodynamics simulation of the Sco-Cen superbubble in a 
	homogeneous environment.
	 Columns from left to right show, respectively, sightline-averaged logarithmic gas density, 
	 logarithmic temperature in the midplane, 
	pressure in the midplane, and current, sightline-averaged
	mass density in \al, all in cgs units.
	UCL is initiated at time $= 0$ in the centre of the computational domain. LCC is formed at 
	2~Myr outside the UCL bubble,
	towards its lower left. At 5~Myr, the interface between the individual bubbles is eroding (top row).
	USco forms inside the UCL-LCC superbubble towards the upper right at 10~Myr. 
	The expanding wind of its 60~\msun star
	is easily identified in the pressure plot at 10.85~Myr (second row). 
	The overall pressure in the superbubble is high at this time
	due to two supernovae in UCL exploding at 10.48~Myr and one in LCC at 10.61~Myr. 
	The massive star in USco explodes at 14.86~Myr. At 14.90~Myr
	(third row), it pressurises the superbubble again via a \alu-rich shock wave. 
	At the end of the simulation (15.80~Myr, bottom row, representing the present epoch),
	\al is uniformly distributed throughout the superbubble. 
	The pressure is low, but the temperature is still high enough
	to expect X-ray emission throughout the superbubble.
	}
   \label{fig:sim}%
\end{figure*}

\newcommand{\judims}{./}
\setlength{\flength}{0.33\textwidth}
\begin{figure*}
  \centering
 	\includegraphics[angle=-0,width=\flength]{\judims 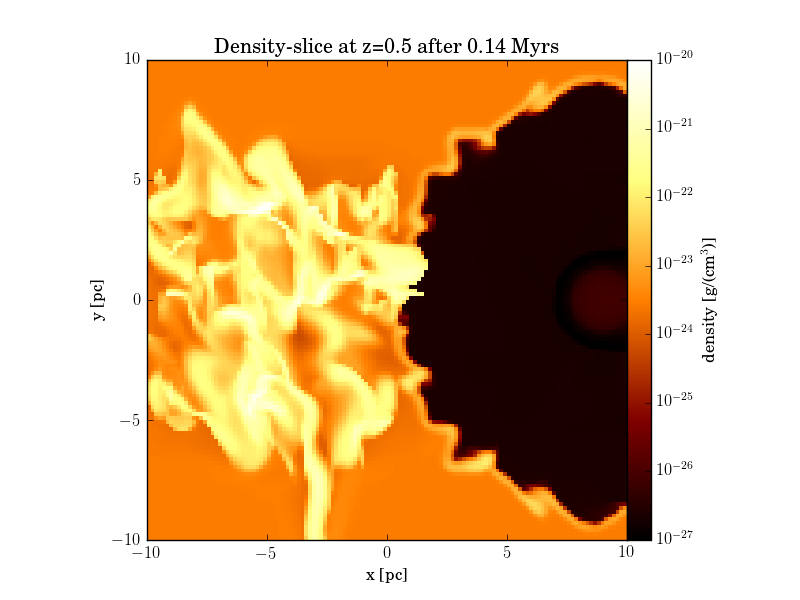}
  	\includegraphics[angle=-0,width=\flength]{\judims 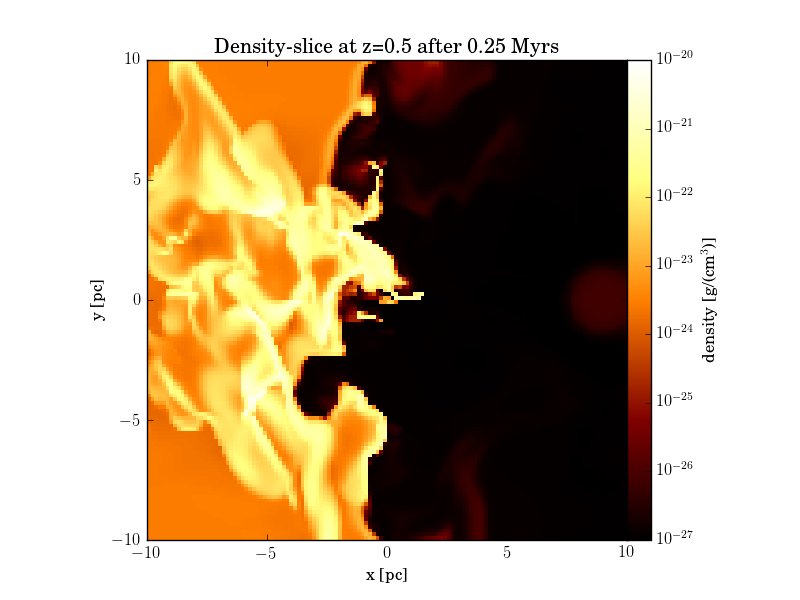}
 	\includegraphics[angle=-0,width=\flength]{\judims 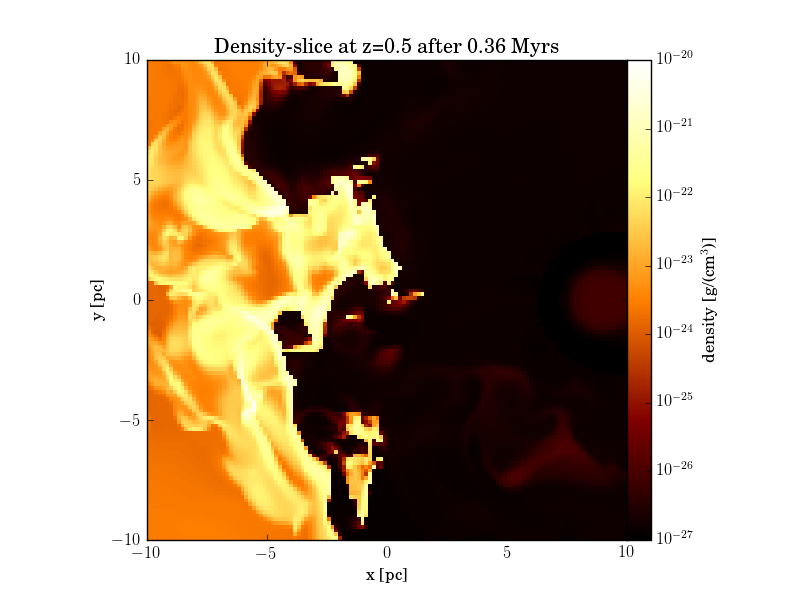}
 	\includegraphics[angle=-0,width=\flength]{\judims 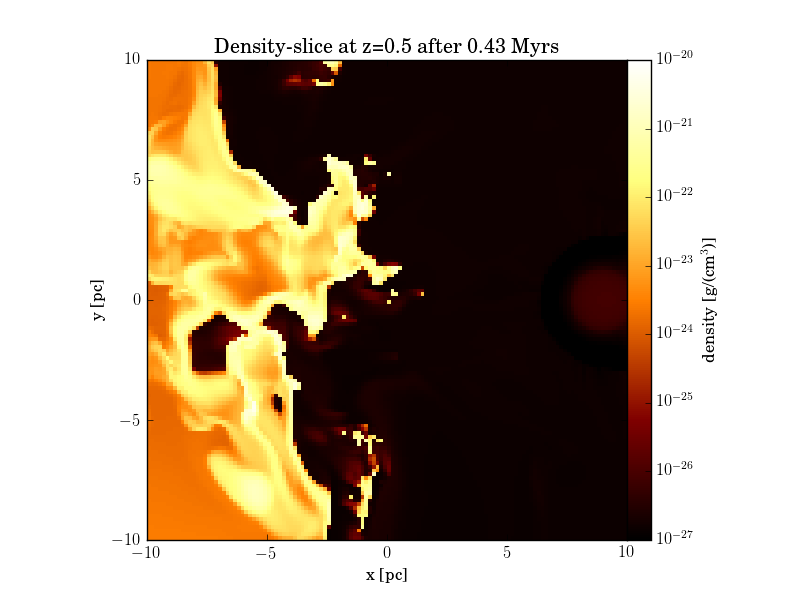}
  	\includegraphics[angle=-0,width=\flength]{\judims 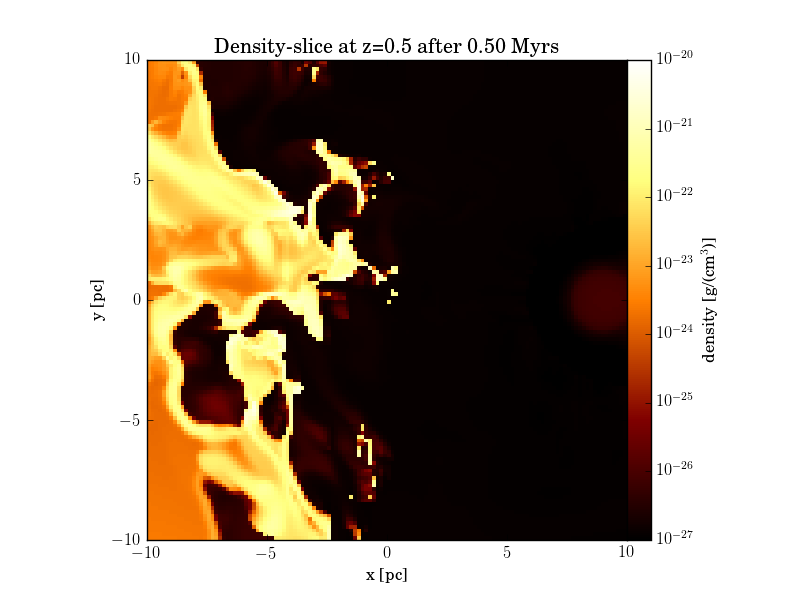}
 	\includegraphics[angle=-0,width=\flength]{\judims 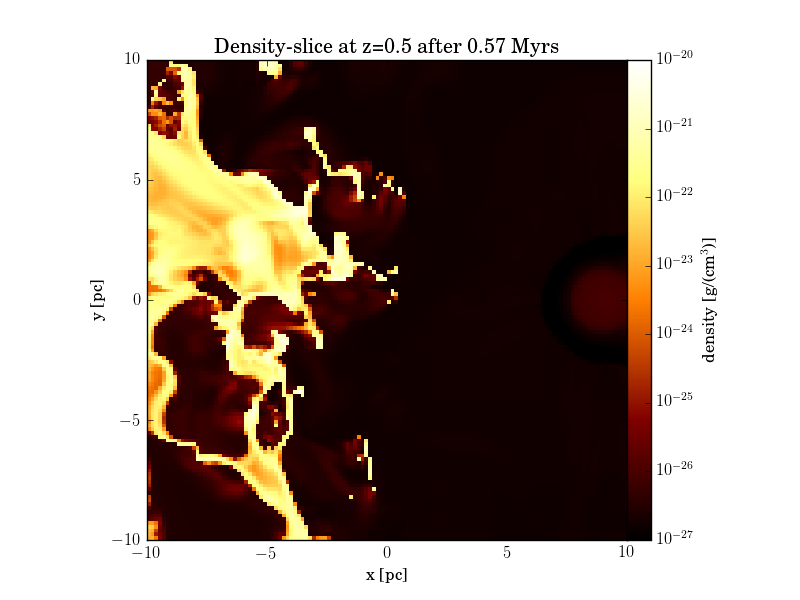}
 	\includegraphics[angle=-0,width=\flength]{\judims 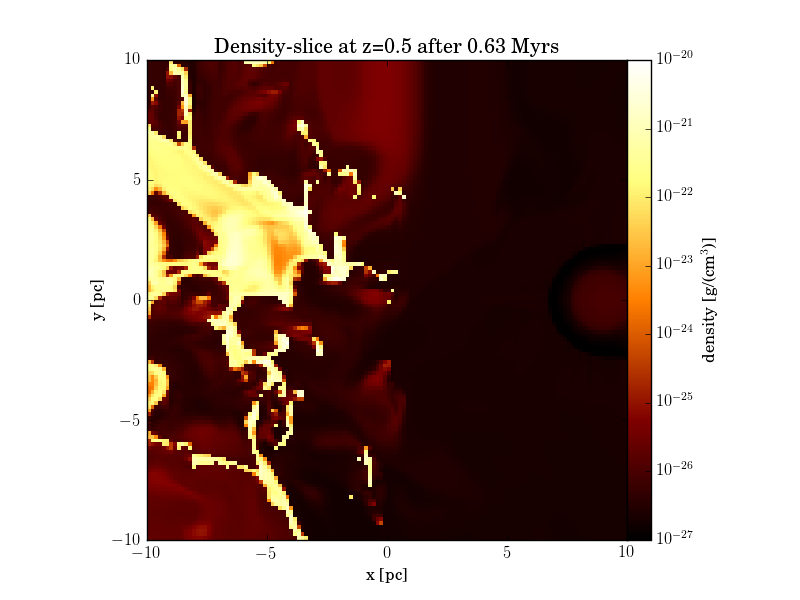}
  	\includegraphics[angle=-0,width=\flength]{\judims 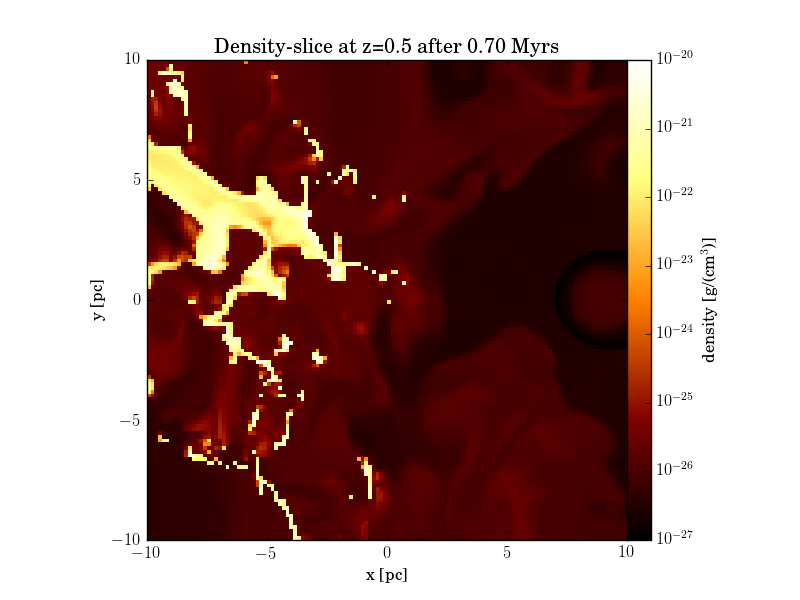}
 	\includegraphics[angle=-0,width=\flength]{\judims 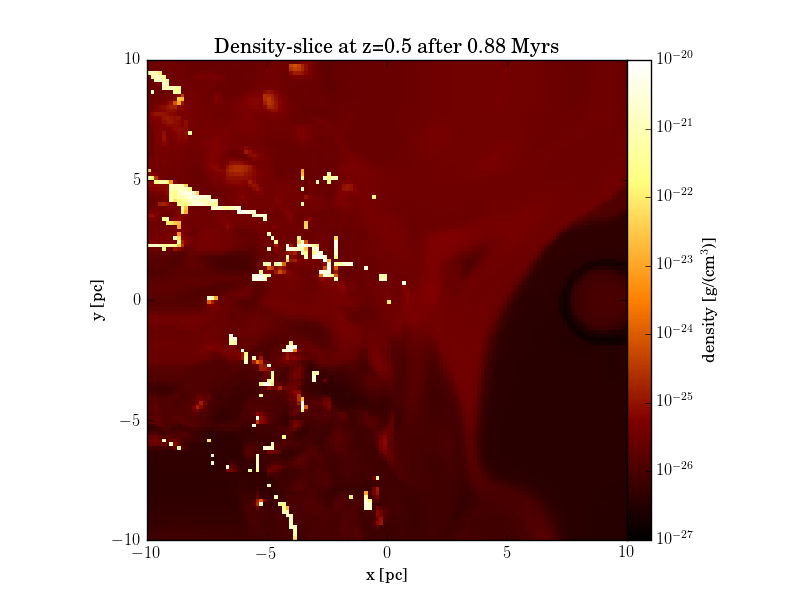}
\caption{Time evolution of a 3D hydrodynamics simulation of a turbulent cloud
	overrun by a superbubble. Shown are midplane slices of the density with a 
	logarithmic colour scale at different times, as indicated in the individual panels.
	The dense cloud first expands due to its inherited turbulence. The shock and 
	supershell have little impact on the dense regions of the clouds. The hot, gas,
	however penetrates the low density parts of the cloud, thus surrounding the denser
	parts. The intermediate
	density regions are thus pushed into the dense clumps by the overpressure of the hot gas. 
	About 40~per cent of the initial cloud mass is squashed into distributed dense clumps in this way, 
	after about 0.5~Myr. The rest has been swept away.
	A movie is provided with the online version.
	}
   \label{fig:simc}%
\end{figure*}
\subsection{Ejecta from massive stars}\label{sec:26al}
Stellar winds and supernova explosions return stellar material back
into the surrounding interstellar medium. Such ejecta are enriched 
with products of nuclear reactions, i.e., with new nuclei that were 
not part of the composition of gas that formed those stars. Among 
those new nuclei are radioactive nuclei, whose decay provides a direct 
observational signature of massive star feedback. 
The \al  isotope with a radioactive lifetime of 1.04 Myrs 
and emission at 1809~keV has been observed as such 
proof of current-generation massive-star nucleosynthesis 
\citep{Mahonea82,Diehlea95, PD96}
\al  ejection from a coeval group of stars is expected to 
commence after a few Myr 
from the winds of the most-massive (Wolf-Rayet) stars, and continue 
from less massive stars and their core collapse supernovae until
$\approx 30-40$ Myrs after the star forming event 
\citep[for details see the population synthesis study of][]{Vossea09}.

INTEGRAL with its gamma-ray spectrometer instrument SPI
\citep{Vedrea03} has collected significant observational data from
the larger Sco-Cen region since its launch in~2002, and during the
extended INTEGRAL mission. The total exposure in this region amounts
to about 10~Msec, with variations of about a factor ten across the
larger Sco-Cen region. 

From earlier analysis of about 6~Msec of exposure after 5~mission
years, the \al  signal from the Sco-Cen region could be detected and
discriminated from the Galaxy's bright \al  gamma ray emission
\citep{Diehlea10}. Defining the location of \alu -rich ejecta is the main
uncertainty in such analysis; the signal surface brightness is too
faint to allow imaging from the data themselves at the required
precision \citep[see, e.g.,][]{BJR15}. Assuming that the Upper Sco group
of stars is the most plausible source of any current radioactive \al
from this region (cmp. Table~\ref{t:sc-overview}), 
the \al  signal was shown to be of
diffuse nature and extended by about $10\deg$ in radius around
$(l,b)=(350\deg,20\deg)$, which corresponds to about 25~pc at the distance of
USco. On the other hand, if ejecta were to stream freely into
surroundings for one \al  lifetime at typical velocities within a
tenuous superbubble of $\approx 300$~\kms, the \alu -filled region could be
the entire Sco-Cen superbubble 
\citep[also compare the simulation below]{Kretschea13,Krausea15a}. Hence, we may assume that these gamma-ray
observations only saw a fraction of the \al  that is located near its
sources and with highest surface brightness, while the \al  emission
from more distant regions fades and becomes confused with the bright
emission of the Galaxy. Still, the flux of \al  gamma rays found with
INTEGRAL in this way from USco is 
$(6.1\pm1.0) \times 10^{-5}$~ph~cm$^{-2}$~s$^{-1}$,  and
corresponds to $1.1 \times10^{-4}$~\msun~of \al  \citep{Diehlea10}.  In a recent
re-analysis from 13 years of observations, this is confirmed: For a
slightly larger region around USco of $16\deg$ radius
around $(l,b)=(340\deg,23\deg)$, a flux of 
$(7.6\pm1.4) \times10^{-5}$~ph~cm$^{-2}$~s$^{-1}$
-- corresponding to $(1.3\pm0.3) \times10^{-4}$~\msun~of \al --
was determined; this comprises a $6 \sigma$ detection
of \al  emission from USco \citep[Fig.~\ref{fig:26Al}, ][ Siegert et al., in preparation]{Siegert2017_PhD}. 
We note that the new analysis also resulted in continuum level
towards Sco-Cen below the average over the sky, and hence a negative value
in Fig.~\ref{fig:26Al}. The \al measurement is relative to the continuum level and 
therefore not affected by this offset.
Typical massive star yields of \al  in models from Wolf Rayet stars
and core collapse supernovae are of the order of  $10^{-5}-10^{-4}$~\msun
~\citep[e.g.,][]{Diehlea06,CL13}, so that the observed \al is
consistent with contributions from several massive stars. 

Gamma ray spectroscopy with the \al  line has obtained a precision
level that allows derivation of kinematic constraints on the
radioactive ejecta \citep[e.g.,][]{Kretschea13}: The spectral
resolution of the SPI Ge spectrometer allows to resolve the line, and
determine its Doppler broadening from irregular motion of the ejecta
down to $\approx 100$~\kms. Bulk velocity constraints can be derived from the
line centroid, comparing to the laboratory energy of \al  decay 
\citep[1808.73~keV][]{Endt98}. An application to the signal obtained for
USco indicated a bulk flow towards us at
$(137 \pm 75)$~\kms \citep{Diehlea10}.
Our re-analysis from 13 years of INTEGRAL
data confirms this, with slightly better precision, at 
$(154 \pm 53)$~\kms. 
Any Doppler broadening of the line is constrained to 
below 150~\kms ($2\sigma$) \citep[ Siegert et al., in preparation]{Siegert2017_PhD}.

This confirms the impression from the \HI data above 
(compare Fig.~\ref{fig:HI-USco}) that the sheet of dense gas that 
separates the local bubble from the Scorpius-Centaurus superbubble 
is currently eroded by the hot gas that is 
being pushed towards us into the local bubble.

\newlength{\swidth} 
\setlength{\swidth}{0.33\textwidth}
\begin{figure*}
  \centering
  \includegraphics[angle=0,width=\swidth]{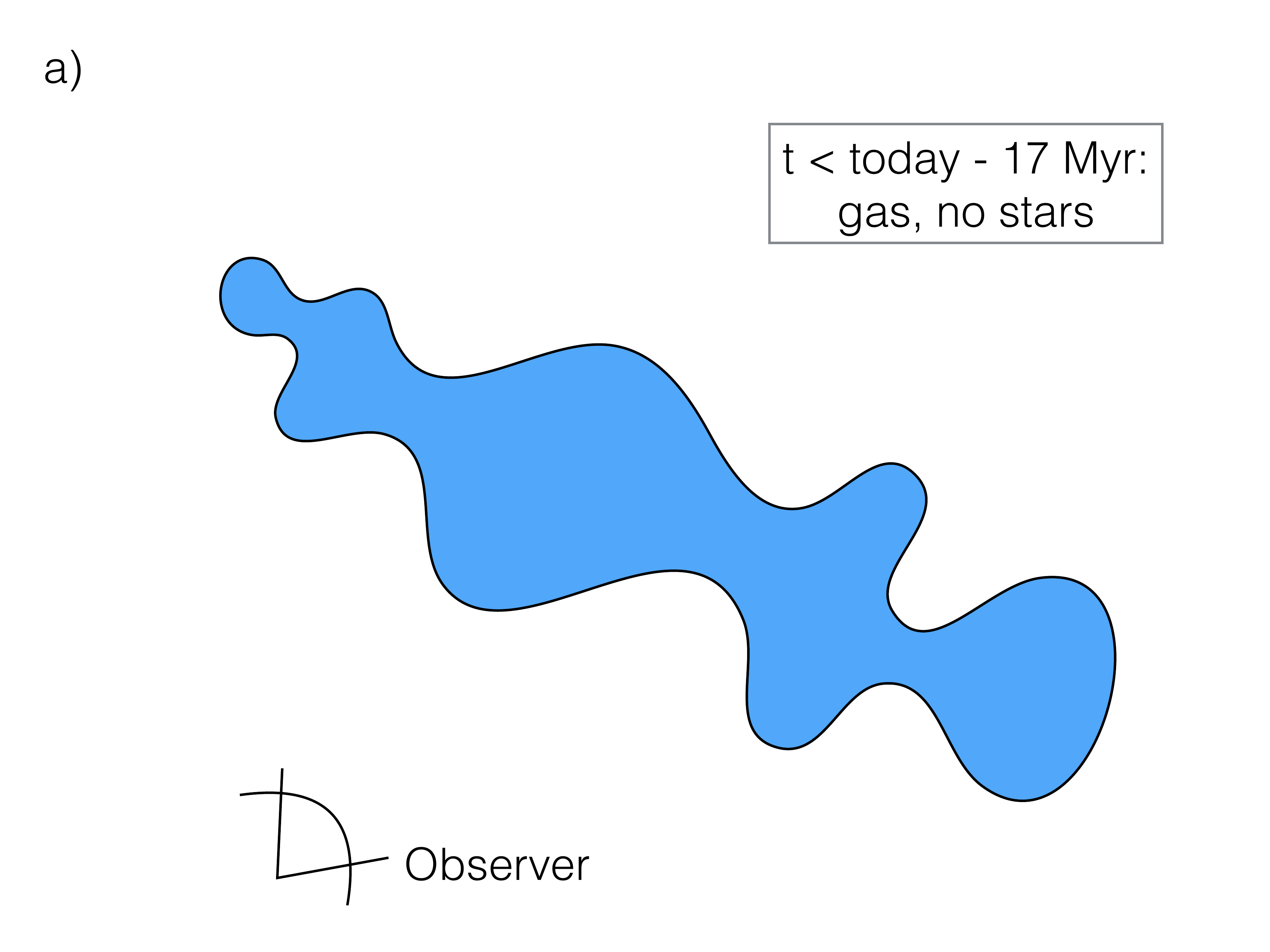}
  \includegraphics[angle=0,width=\swidth]{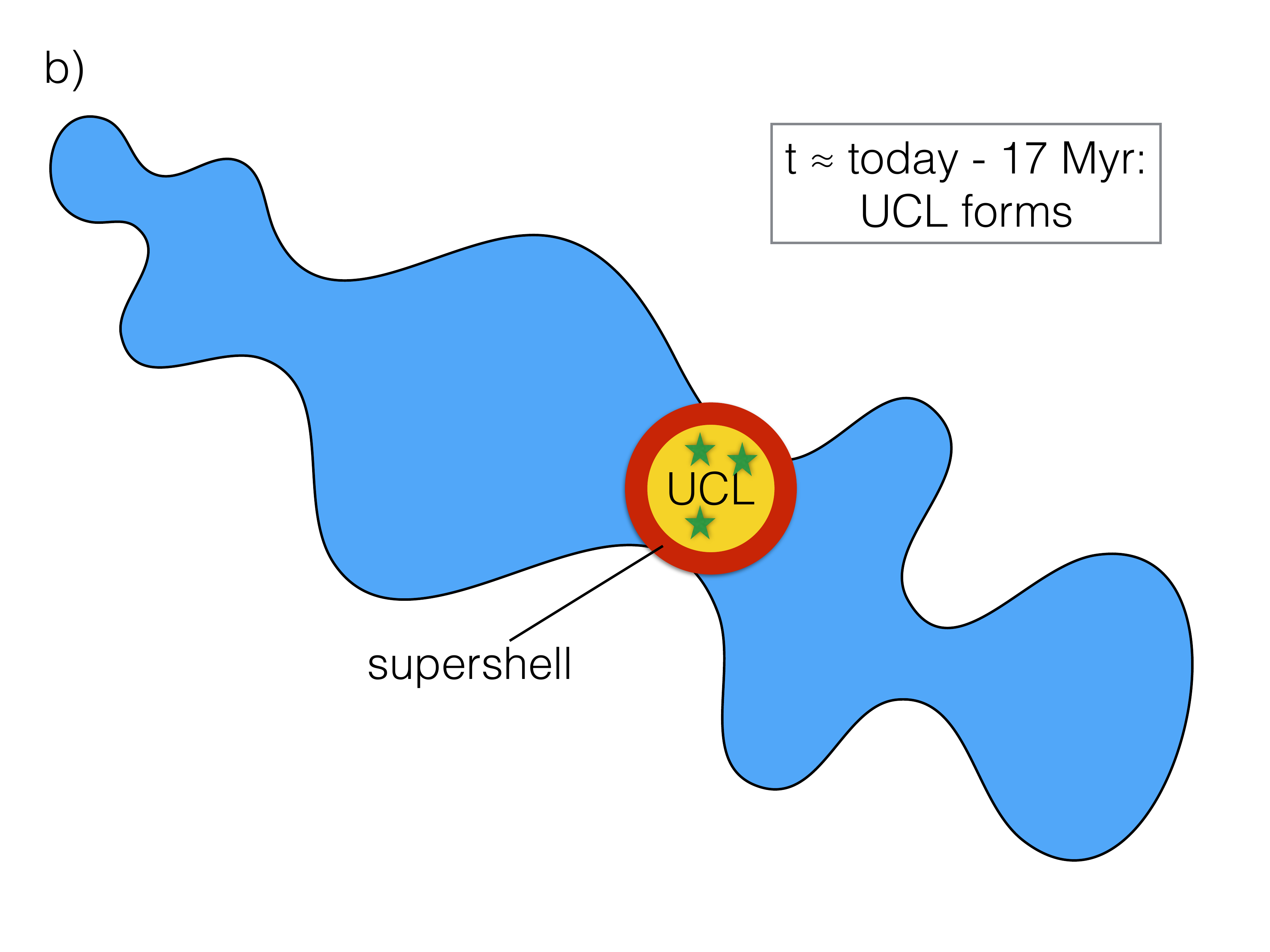}
  \includegraphics[angle=0,width=\swidth]{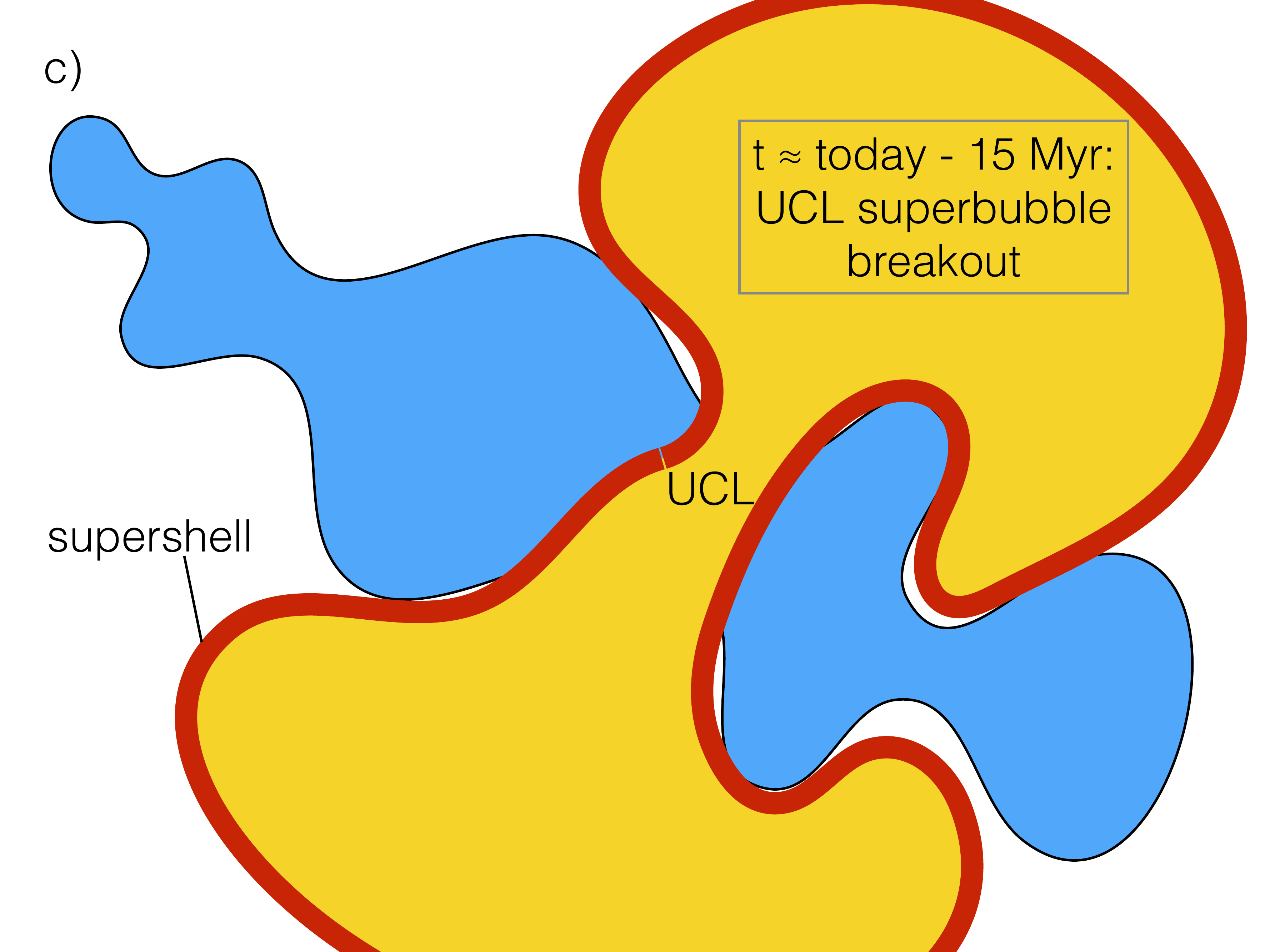}
  \includegraphics[angle=0,width=\swidth]{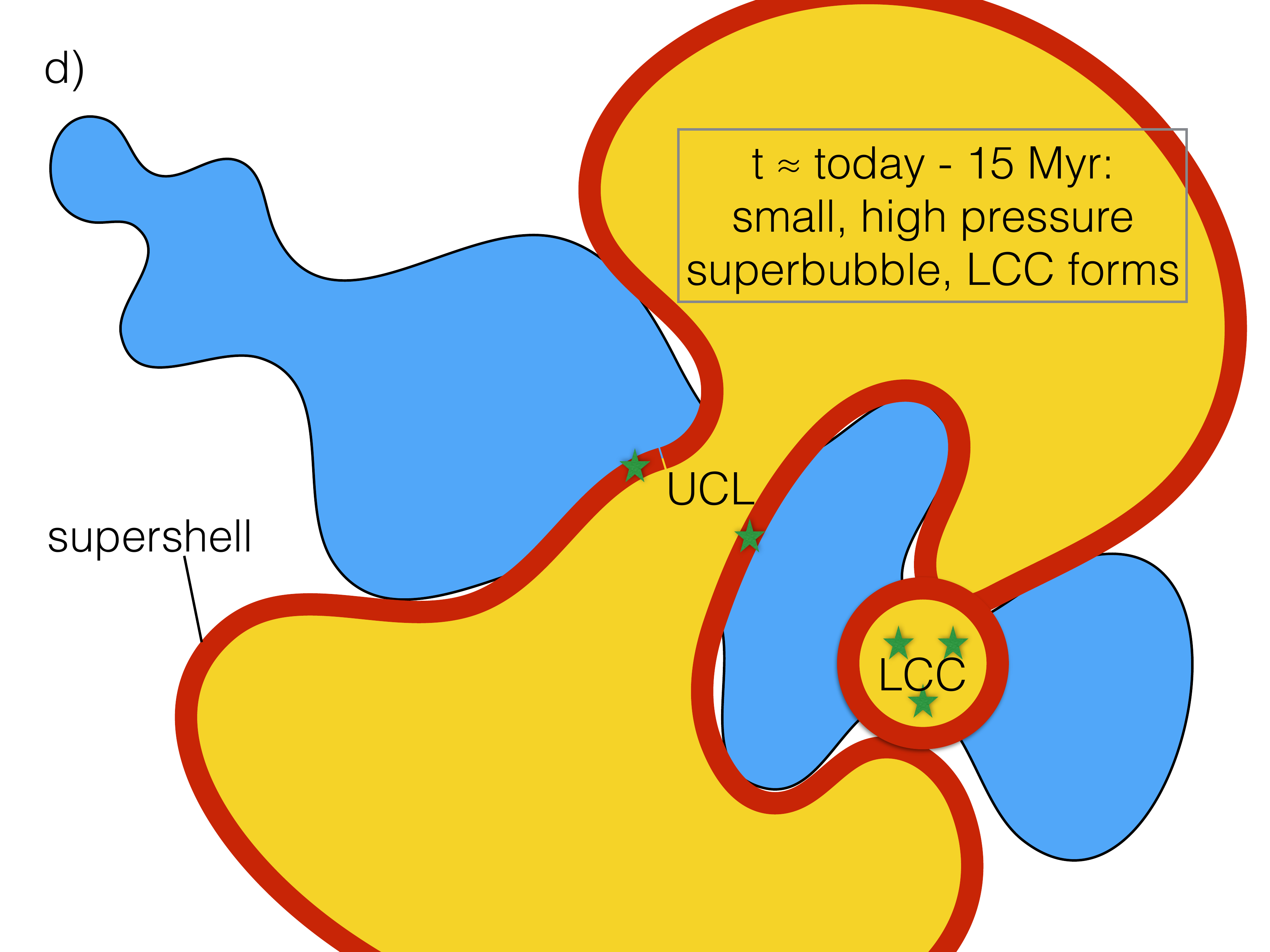}
  \includegraphics[angle=0,width=\swidth]{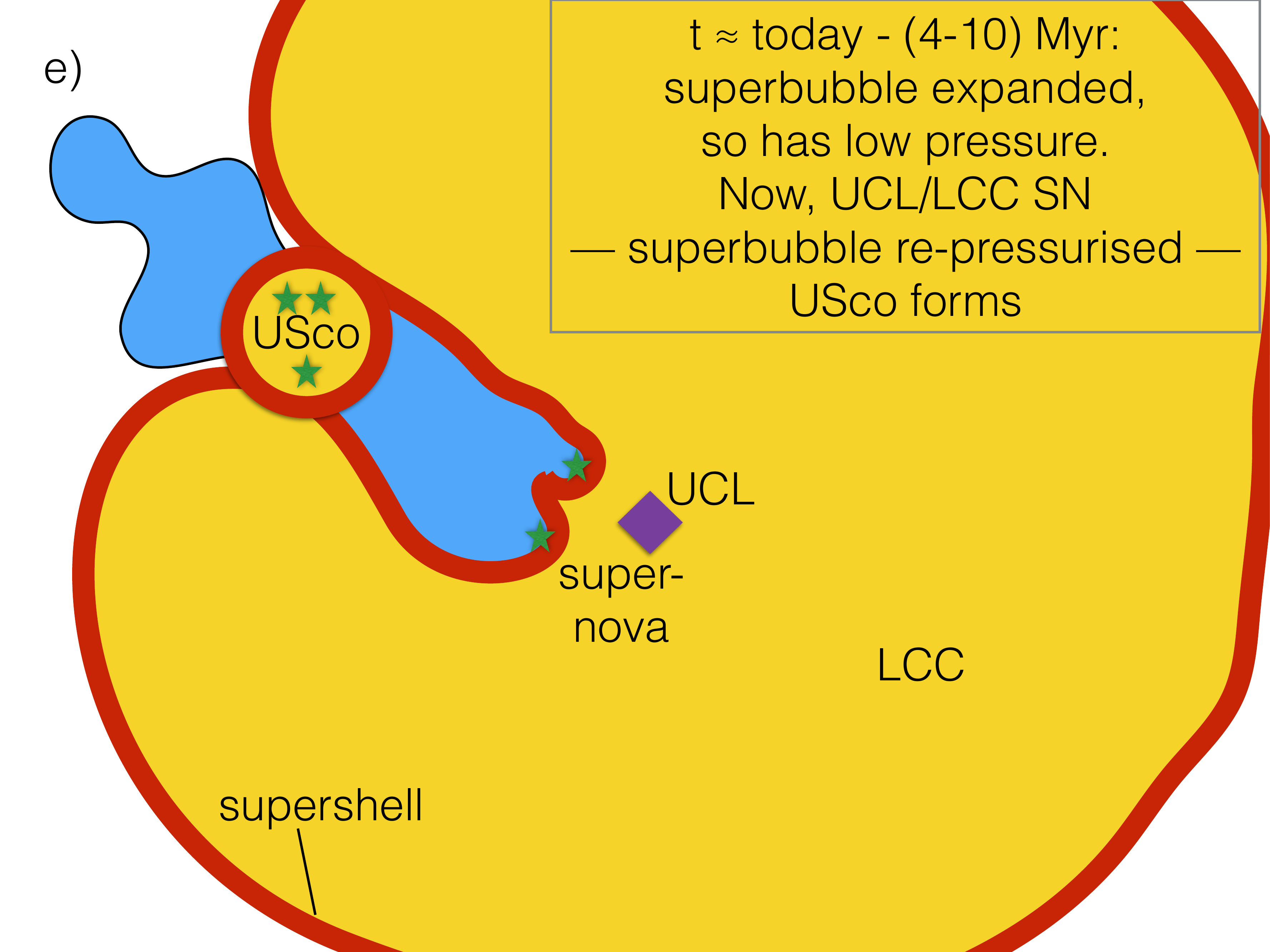}
  \includegraphics[angle=0,width=\swidth]{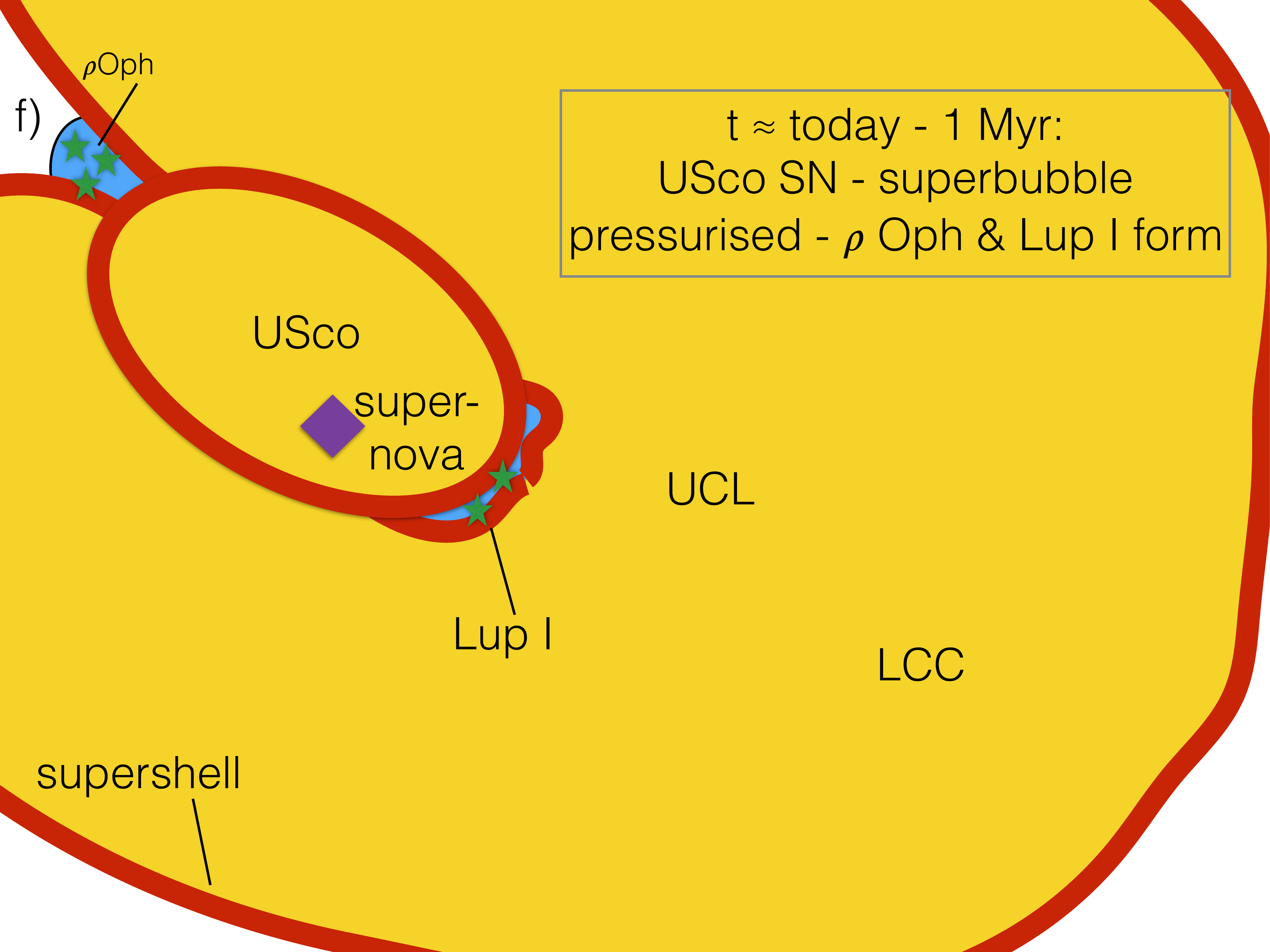}
  \caption{Sketch of evolutionary scenario on the basis of the results of the present study.
	a) Before 17~Myr ago the elongated gas cloud in Sco-Cen has not formed any stars yet. 
	b) 17~Myr ago: Something causes the onset of star formation in UCL. A superbubble forms.
	c) 15~Myr ago: The superbubble has expanded beyond the scale height of the gas stream. It now
	expands quickly into more tenuous regions, but cannot penetrate into the dense clumps in the parent cloud.
	d) 15~Myr ago: Star formation starts in LCC. This could be spontaneous, or enabled
	by compression by the UCL superbubble or a supernova in a more distant, older member
	of the elongated cloud / star cluster system.
	e) 4-10~Myr ago: supernovae pressurise the superbubble and squash the surrounded part of the parent cloud.
	Star formation sets in at USco. Winds from USco then sweep up the rest of the cloud.
	f) 1~Myr ago: A supernova in USco pressurises the superbubble and its rim once again.
	Rho Ophiuchus and Lupus~I start star formation.
	}
   \label{fig:sketch}%
\end{figure*}

\subsection{Hydrodynamic simulations of the superbubble}\label{s:sim}

We used the 3D mesh-refining hydrodynamics code RAMSES \citep{Tey02} to simulate
the Sco-Cen superbubble. 
\hi{We first performed a global simulation of the superbubble and then studied the reaction
of a molecular cloud to the sudden pressure increase when it is overrun by a superbubble.
We  used a standard cooling-heating function taking into account atomic and molecular transitions.
Massive star feedback is simulated by a spherical source region of 1~pc radius with
mass, energy and \al  
input rates as determined from stellar models \citep{Fierlea16a}. 
\subsubsection{Global simulation}
We run the global simulations in a smooth background medium.}
As sources we used a plausible configuration of 
massive stars (Table~\ref{table:simstars}) given the observed stellar population (Table~\ref{t:sc-overview}).
%
\begin{table}
\caption{Stars used in the the 3D hydrodynamic simulation}             
\label{table:simstars}      
\centering                          
\begin{tabular}{l c c c r}        
\hline\hline                 
Sub\tablefootmark{a} &  $T_\mathrm{f}$\tablefootmark{b}  & $(X,Y,Z)$\tablefootmark{c} & $M_*$\tablefootmark{d} & $T_\mathrm{SN}$\tablefootmark{e}  \\    
\hline                        
   UCL      &  0                        & -2.27, 3.56, -3.89 &        60            &   4.86\\      
   UCL      &  0                        & -2.27, 3.56, -3.89 &        32            &   7.22\\      
   UCL      &  0                        & -2.27, 3.56, -3.89 &        25            &   8.61\\      
   UCL      &  0                        & -2.27, 3.56, -3.89 &        20            & 10.48\\      
   UCL      &  0                        & -2.27, 3.56, -3.89 &        20            & 10.48\\      
   UCL      &  0                        & -2.27, 3.56, -3.89 &        15            & 15.07\\      
\hline                                   
   LCC      &  2                        & -26.6, -47.6, -25.3 &        60            & 6.86\\      
   LCC      &  2                        & -26.6, -47.6, -25.3 &        25            & 10.61\\      
   LCC      &  2                        & -26.6, -47.6, -25.3 &        15            & 17.07\\      
\hline                                   
   USco      &  10                     & 50.6, 27.5, 6.16 &        60            & 14.86\\      
\hline                                   
\end{tabular}
\tablefoot{ 
\tablefoottext{a}{Subgroup name.}
\tablefoottext{b}{Formation time in Myr.}
\tablefoottext{c}{Fixed position of the respective star on the grid in pc.}
\tablefoottext{d}{Mass of the star in solar masses.}
\tablefoottext{d}{Explosion time of the star in Myr.}
}
\end{table}
Due to stochasticity of the IMF, the number of past supernovae is highly uncertain.
To be conservative, we decided for a number of massive stars on the low side of the 
literature estimates (compare Table~\ref{t:sc-overview}). 
The initial density distribution was homogeneous at $8.5 \times 10^{-24}$~g~cm$^{-3}$, 
consistent with the total \HI mass we find in
the USco loop \citep[$10^4$ \msun,][]{Gaczkea17} and in the wider region of 368,000~\msun ~\citep{Poepea10}. 
The initial equilibrium temperature was 1647~K and the box size $(290\mathrm{pc})^3$.
Here, we do not attempt to include
dense gas clouds. \hi{This simulation} can therefore not address nested shells or details of late star formation directly.
Current observations should be compared to the final state of the simulation at $t=15.8$~Myr.

We show the time evolution of the simulation up to the present epoch in Fig.~\ref{fig:sim}.
UCL and LCC form here separate superbubbles that merge from about 5~Myr after initialisation of the
simulation (10~Myr ago). The pressure during this period is consistently high, in excess of
10 times that of the ambient gas. 

Around 10~Myr, the superbubble experiences again high pressure due to 
three supernovae in LCC and UCL within about 0.1~Myr. This is of course an arbitrary feature of 
our modelling, but demonstrates that such a coincidence is well possible.
At this time, the superbubble has engulfed the position of USco. 
Any dense clouds in this region
\hi{would thus be compressed.} It is possible that USco stars formed in this way.
In our simulation, feedback from USco starts at this time (Fig.~\ref{fig:sim}, second row).

Pressure in the superbubble then declines until the 60~\msun ~star in USco explodes 
(Fig.~\ref{fig:sim}, third row). The superbubble experiences again overpressures in excess
of a factor of 10. Any remaining clouds, possibly swept up by the wind of the 60~\msun ~star,
would again be compressed and potentially prompted to form stars. This would be consistent 
with the recent and approximately coeval onset of star formation in the $\rho$~Ophiuchus 
and Lupus~I molecular clouds.

We end the simulation 1~Myr after the supernova in USco (Fig.~\ref{fig:sim}, bottom row). The superbubble 
has now a diameter of more than 200~pc, well consistent with observations. Its shape
still bears a memory of its formation history, as is the case for the real, elongated Sco-Cen superbubble.

We predict the current superbubble temperature to be spatially variable between $10^6$ and $10^8$~K,
well in the regime accessible by X-ray observations. This is consistent with the \rosat detection.

The snapshots in Fig.~\ref{fig:sim} show phases of particularly high \al abundance. This varies by orders 
of magnitude during the evolution of the superbubble. During times of actively blowing massive star winds, 
\al can be restricted to less than half of the superbubble volume. For the present time, 1~Myr after the last massive star 
($> 15$~\msun) explosion, we predict a uniform \al distribution. This is consistent with the updated, 
now more extended \al detection presented in Sect.~\ref{sec:26al}.

\hi{
\subsubsection{Surrounded and squashed cloud}\label{ssc}
In order to study the effect of a superbubble engulfing a dense cloud,
we first produced a turbulent cloud in a periodic box, and then put it on 
a RAMSES grid next to a massive-star wind.}

\hi{
For the turbulent cloud setup, we followed the procedure described in \citet{MB15}.
Isothermal turbulence was driven in a $1024^3$ periodic box using the ATHENA code 
\citep{GS05,GS08,Stonea08,SG09}. Smaller scale turbulence developed
self-consistently from the larger-scale driving. When the turbulence was well developed,
we scaled the box to 10~pc on a side, a mean number density of 100~cm$^{-3}$
(giving a total mass of $\approx 5700 M_\odot$) and a temperature of 10~K.
The rms-velocity of the cloud is 18~km~s$^{-1}$. The cloud
has time to expand initially on the RAMSES grid and loses its memory of the cubic setup
to some degree
before significant interaction occurs.
}

\hi{
The cloud was imported on a 3D Cartesian RAMSES grid with 20~pc on a side resolved by 128
cells. Self-gravity and adaptive mesh refinement was not used. 
Boundary conditions are open so that the flow may leave the box freely.  
The grid was initialised with a density of $2.9\times 10^{-24}$~g~cm$^{-3}$,
adding up to a total mass of $\approx 300 M_\odot$ for the whole box.
The massive star was then positioned
8~pc away from the cloud surface on one of the symmetry axes. We used the parameters for the 
$60 M_\odot$ rotating star in the models of \citet{Ekstrea12a} with wind properties as 
in \citep{Fierlea16a}.
}

\hi{
Snapshots of the logarithmic density distribution in the midplane of the simulation are 
shown in Fig.~\ref{fig:simc}. The massive star wind produces a typical superbubble 
structure, i.e. a low density bubble with a cool and dense shell that clumps
due to the Vishniac instability. The superbubble starts to interact with the turbulent cloud 
about 0.14~Myr after the start of the simulation. Subsequently, the hot gas penetrates
the low-density parts of the cloud, and surrounds the higher density parts.
From 0.43~Myr to 0.57~Myr, individual low-density regions within the cloud grow at the
expense of the denser parts.
Until 0.88~Myr, the cloud is completely compressed into a large number of resolution-scale
fragments. 
While all parts of the clouds show significant motion at the start of the interaction, they have
slowed down to the velocity of the hot bubble gas (0 in the frame of the simulation)
towards the end by interaction with the hot gas.
The typical mass of the fragments is a fraction of a solar mass.
The total mass in the box at the end of the simulation is $2353~M_\odot$, which is almost
entirely in the cold fragments.
}

\hi{If the fragments in our simulation represented the formation sites of individual stars,
the mechanism would coordinate star formation in a volume of 10-20~pc diameter within
about 0.5~Myr, much faster than the sound travel time in molecular gas for this
length scale, which would be $>10$~Myr. This is similar to the length and timescales deduced
from observations of the stars in the Sco-Cen subgroups \citep[compare][]{PreibMam08}.
}

\section{Discussion: Deciphering the interstellar medium in Sco-Cen}\label{sec:itp}

\citep{PreibMam08} present the case for triggered star formation in USco: With a stellar velocity 
dispersion of $\approx 1.3$~\kms, an initial size of the stellar subgroup of 25~pc and an age spread
for the stars of 1-2~Myr, the formation of the stars cannot possibly have been coordinated
by internal processes in the parent cloud. An external agent associated with a characteristic
velocity of at least 20~\kms~is required. The timescale is consistent with a shock wave driven by 
massive star feedback in UCL that induced collapse in an existing molecular cloud. 

A problem of this scenario
is that any molecular clouds in the region could have formed stars 
also without a shock wave passing. It would therefore be unclear why star formation was not proceeding
independently in such clouds for several Myr. Further, if the clouds were indeed present for several Myr, why did 
the shock wave that triggered star formation in USco not trigger star formation in Lupus~I, about 
half way to USco?

Clues might come from the results shown here, summarised in the sketch in Fig~\ref{fig:sketch}: 
In a linearly extended stream of gas, as suggested by the morphology and arrangement of the 
stars in Sco-Cen OB2, as well as a sequence of progressively older star clusters \citep{BouyAlves15},
superbubbles produced by massive stars will more easily break out towards the short 
axis of the parent cloud, similar
to larger superbubbles leaving the disc of the Milky Way \citep[compare][]{MacLowea89,Fierlea12a}.
These superbubbles would then engulf some of the until then more diffuse parts of the cloud.
We suggest that we see this currently in USco. The \HI observations show a loop with constant diameter
in adjacent velocity bins. It is therefore likely that this superbubble broke out of the parent cloud,
now engulfing its remaining part,  the $\rho$ Ophiuchus and the Lupus~I cloud. 

Superbubbles are a very important element for understanding the system: The sound speed
in superbubbles is high, varying between a few 100 and a few 1000~km/s. This is seen
in our simulation (Sect.~\ref{s:sim}) and evidenced by the X-ray measurements in Sect.~\ref{s:xray}. 
Fast communication between remote parts of the cloud thus becomes possible.

Supernovae increase the pressure in the system by typically by a factor of 10 or more, every few Myr.
More rarely, supernovae will occur within a short enough time interval 
\citep[$\lesssim$ few $10^5$ yrs, e.g.,][]{Krausea13a} that the pressures add up. For dense
gas where the effective index $\gamma$ for a polytropic equation of state, $p\propto \rho^\gamma$,
is close to one, this means that cloud densities have to increase also by a factor of about ten.

\hi{Our simulation of a cloud initially extended over $\gtrsim10$~pc when starting to interact with a superbubble
confirms this hypothesis: The low density regions of the turbulent cloud are penetrated by the hot bubble gas,
and the intermediate density regions are then pushed into the high density clumps, so that a system of low-dispersion,
widely spread out clumps occurs that may form stars in a coordinated fashion. 
About 40 per cent of the mass present in the initial setup has thus fragmented into an
ensemble of dense clumps with low velocity dispersion. If these clumps where the seeds of a stellar population,
their low velocity dispersion and spatial extent would agree well with observations of the Sco-Cen subgroups,
UCL, LCC and USco,
and other OB~associations \citep[compare][Sect.~6.1]{PreibMam08}. 
This would solve the problem how to coordinate the onset of star formation over a distance of more than 
10~pc, where the velocity dispersion of the stars is only $\approx 1$~km~s$^{-1}$.
The simulation is of course highly idealised. We do not include self-gravity, which would in reality make the 
cloud contracting even before the hydrodynamic effects have time to act. Had we, however, included
gravity, we would not have been able to demonstrate that the hydrodynamic effects alone are 
sufficient and possibly decisive to produce such an ensemble of distributedly formed stars.
Because of the lack of gravity and magnetic fields in the simulation, we do not attempt to constrain the
mass spectrum of the clumps any further at this stage.}

\hi{Both of our simulations neglect magnetic fields. Comparison of simulations of the larger-scale
interstellar medium with and without magnetic fields show that the magnetic field generally does not
impede superbubble expansion \citep{dAB04,dAB05}. Magnetic fields are important in dense gas, where
they are expected to provide dynamical support \citep[e.g.,][]{dAB05,II08,II12,KB15,Valdivea16} and 
can potentially suppress clumping instabilities \citep{Vish83,Ntormea17}.
This may well set the actual size scales of the resulting clumps in our simulation that included the cloud.
We would also expect that the actual star formation sites are filamentary, as 
the dense regions in the simulations of \citet{II12}, and as seen in recent observations of 
\citet{Pokhrea18}. We focus here, however, on the connection to the larger scale, and refer to these
other studies for details of the evolution on smaller scales. While magnetic fields have in principle
the potential to suppress the clumping instability in supershells, this is not required to
explain the HI observations of the USco supershell, where the filamentary appearance of the shell
suggests that the Vishniac instability is acting. Magnetic fields might, however, explain the
absence of molecular gas in the HI shell around USco. We stress that the Vishniac instability is not crucial 
for our scenario. Suppression of instabilities at the shell surface of our global simulation would reduce
the extent of the mixing zone, which is responsible for most of the radiative cooling in the bubble.
This would therefore increase the size of the simulated bubble, such that 
a smaller number of massive stars would be required to reach the observed extent.
}

\hi{The blow-out of the HI supershell around USco suggests that Sco-Cen formed from a large 
sheet of dense gas, similar to the dense sheets in large-scale simulations of the interstellar medium
\citep[e.g.,][]{dAB04,dAB05,Seifrea17}. In these simulations these are "walls" of several hundred 
parsec sized superbubbles. \citep{Seifrea17} show that molecular gas builds up in such sheets
over millions of years. \citep{DawsonJea15} may have observed this process at the interface of two HI supershell 
in the Carina arm of the Milky Way. Giant molecular clouds formed in this way are less violently formed
\rvb{than} in typical colliding flow simulations of molecular cloud formation \citep[e.g.][]{II08,KB15}, 
because the pressure in the surrounding
superbubbles is modulated by the actual energy production of the massive stars. 
This might make the clouds comparatively tenuous and delay the formation of stars
\citep{DawsonJea15}. A more gradual buildup of molecular clouds is also seen in more sophisticated, 
multi-phase colliding flow simulations \citep{II12}. \citet{Inutsea15} detail a scenario
of magnetised molecular cloud and star formation along these lines. In general,
molecular cloud formation at the interface of HI supershells is frequent \citep{DawsonJea13}.
Near the solar circle, molecular cloud formation is
only weakly coupled to spiral arm passages \citep[e.g.][]{Kodea16}.
}

\hi{How could the actively star-forming Lupus~I cloud have formed in this scenario?}
Our analysis of multi-wavelength observations of the Lupus~I molecular 
cloud resulted in a double peaked column density PDF
which we interpreted as evidence for compression of part of the cloud, possibly by the most recent 
supernova shock wave having originated in USco \citep{Gaczkea15}. The two peaks differ by a factor 
of about three, and in the slightly denser, southern part of the cloud this was obviously enough to trigger
star formation. 

\hi{The gas that the Lupus~I molecular cloud now consists of was probably the most tenuous part of the dense 
gas sheet out of which the OB association formed. Otherwise, it would not be the last part to form stars.
It is therefore likely that first, the superbubble initially formed around UCL and then the one around USco, of which
we still see the HI shell,
have contributed to sweep up the gas. This has likely resulted in a denser, more compact cloud,
comparable to the current extent of Lupus~I. The supernova in USco 1~Myr ago could then have pressurised
once more the superbubble. Now parts of the Lupus~I cloud were finally dense enough, the hot gas penetrated
the tenuous parts of the cloud, surrounded the denser regions and pushed the less dense ones into them, as
shown in our simulation, and thus triggered the formation of stars. A similar scenario could apply to the nearby 
$\rho$~Ophiuchus molecular cloud.
}

The natural timescale between major star formation events for such a "surround \& squash" scenario 
of propagating star formation would be the time from star formation in a region till the massive stars
explode as supernovae, and the typical time in between supernovae. This is a few Myr for the typical
clusters of $\approx 10^3$~\msun~in the region.  \citet{BouyAlves15} find a continuing sequence
of five star forming regions beyond LCC with similar individual masses \citep{Piskea08}. Their ages are within
about 25~Myr, which yields timescales between the triggering of a few Myr, consistent
with the scenario we propose. There is an increased timespan of about 30~Myr between
the LCC and the preceding member in the sequence, \object{IC~2602}. This is, however, about the
maximum expected timespan in our triggering scenario, given by the main sequence lifetime of 8-9~\msun~stars,
the least massive ones that would still produce a supernova.
The timescale is also consistent for triggering of USco by a supernova in UCL or LCC, and triggering of 
star formation in $\rho$ Ophiuchus and Lupus~I by a supernova in USco.
We note that independent analysis traces the neutron star RX~J1856.5-3754 back to the Sco-Cen region, and 
in particular USco \citep{Tetzea11,Mignanea13}. The explosion date is inferred to be about 0.5~Myr ago,
entirely consistent with the aforementioned results.

The process has similarities to the big superbubbles produced around disc galaxies
by a jet from an Active Galactic Nucleus as studied in simulations by \citep{Gaiblea12}. The radio lobe that surrounds the galaxy also squashes the
entire gas in the galactic disc due to its large overpressure and enhances the star formation rate 
of the simulated galaxy.

\section{Conclusions}\label{sec:conc}
We have carried out a multiwavelength analysis of the interstellar medium in the Sco-Cen region.
From cold and tenuous to hot X-ray gas, including freshly injected nucleosynthesis ejecta,
we find consistent evidence for the gradual 
\hi{trans}formation of clouds, star \hi{formation} and 
the \hi{expansion of the} Sco-Cen superbubble. This superbubble 
has at times surrounded denser clouds of gas, and squashed them with the likely result of 
triggering more star formation. At the current time, this "surround \& squash" scenario 
predicts \al uniformly spread out through
the cluster, which we detected with \itgu. The bubble gas should have X-ray temperatures which we 
found in the \rosat all sky survey. We have performed hydrodynamic simulation\hi{s}
that reproduce key features of our observations.

We suggest a refined scenario for the evolution of Sco-Cen: About 15-17~Myr ago star formation has 
started in an elongated, possibly flattened, gas cloud. Stars first formed in a particular 
part of the elongated cloud, now the UCL subgroup. A superbubble then cleared the immediate 
vicinity, but when it broke out, advancement of the supershell into the dense gas 
cloud almost stops, and a hot bubble forms around it. Hot gas also permeates the 
more tenuous parts of the dense cloud 
which are inside the superbubble, but 
leaves the denser parts \hi{in place and squashes them to higher densities}. 
\hi{This may have contributed to form the LCC subgroup.}
Supernovae in 
UCL and LCC then suddenly increase the pressure within the superbubble after several Myr. This 
\hi{squashes} some gas in the elongated cloud, \hi{now} inside the superbubble, 
\hi{so that it} collapse\hi{s} gravitationally and
form\hi{s} stars, thus producing the USco subgroup.
Massive star winds in USco have then swept up the remaining parts of the cloud, leaving
$\rho$ Ophiuchus and Lupus~I at opposite sides.
We find kinematic and morphological evidence of shell instabilities, 
\rvb{as might be expected for} a cloud of swept-up gas. A recent supernova in USco
could then have triggered star formation simultaneously in these two clouds.

\hi{This scenario of hot gas surrounding and squashing the denser parts of clouds to induce
the formation of stars probably makes the strongest prediction in the area of kinematics. The resulting
star groups should be gravitationally unbound and subgroups should have coherent kinematics 
reflecting the local streaming velocity of the hot gas they were embedded in during the squashing.
Since superbubbles can break out anisotropically from a gas distribution and have
spatially and temporally varying pressure distributions (compare Fig.~\ref{fig:sim}), the
kinematically coherent subgroups would be expected to move in different directions, not necessarily 
away from the next older part in the sequence of star formation.
This is similar to the observed kinematics in Sco-Cen \citep{WM18}:
The three subgroups are each unbound and have coherently moving sub-units. The velocity dispersion
is so small that the groups cannot have formed on a much smaller spatial scale as bound clusters
and expanded due to gas expulsion.}

\hi{Sco-Cen OB2 is a typical OB association. Molecular cloud and star formation appear to occur
frequently in dense sheets between superbubbles \citep[e.g.,][]{DawsonJea13,DawsonJea15,Inutsea15,Seifrea17}.
A similar surround and squash mechanism might therefore operate frequently in the Milky Way and
other star-forming galaxies.
}
The 'surround-and-squash' scenario \hi{might even apply to} AGN-related superbubbles affecting star formation
\hi{on galaxy scale}.


\begin{acknowledgements}
 \rvb{This work was supported by funding from Deutsche Forschungsgemeinschaft
under DFG project number  PR 569/10-1  in the context of the
Priority Program 1573  “Physics of the Interstellar Medium”.
Additional support came from funds from the Munich Cluster of Excellence
“Origin and Structure of the Universe”. (www.universe-cluster.de). 
MK thanks the Australian Research Council for support via an Early Career Fellowship, DE130101399.
We thank the anonymous referee for very useful comments that helped
to improve the presentation of these results.
This project has received also funding from the European Union's Horizon 2020 research and innovation programme under the Marie Sklodowska-Curie grant agreement No 664931. }
\end{acknowledgements}

\bibliographystyle{aa}
\bibliography{/Users/mghkrause/texinput/references}
\end{document}